\documentclass[a4paper,11pt]{article}
\pdfoutput=1 

\usepackage{jheppub} 

\usepackage[T1]{fontenc} 

\usepackage{slashed}

\usepackage{subcaption}
\usepackage[svgnames]{xcolor}

\title{\boldmath Worldline approach for spinor fields in manifolds with boundaries}

\author[a]{Lucas Manzo}
\affiliation[a]{Instituto de Física La Plata, CONICET and Universidad Nacional de La Plata,\\ CC 67 (1900) La Plata, Argentina.}

\emailAdd{lucasmanzo@fisica.unlp.edu.ar}

\abstract{The worldline formalism is a useful scheme in Quantum Field Theory which has also become a powerful tool for numerical computations. It is based on the first quantisation of a point-particle whose transition amplitudes correspond to the heat-kernel of the operator of quantum fluctuations of the field theory. However, to study a quantum field theory in a bounded manifold one needs to restrict the path integration domain of the point-particle to a specific subset of worldlines enclosed by those boundaries. In the present article it is shown how to implement this restriction for the case of a spinor field in a two-dimensional curved half-plane under MIT bag boundary conditions, and compute the first few heat-kernel coefficients as a verification of the proposed construction. This construction admits several generalisations.}

\begin{document} 
\maketitle
\flushbottom

\section{Introduction}
\label{section:intro}

The worldline formalism is a method to compute different quantities in Quantum Field Theory, such as effective actions, amplitudes, anomalies and partition functions. Unlike the traditional procedure involving Feynman diagrams, this formalism is characterised by the introduction of a point-particle whose dynamics is described using a first quantisation scheme. The transition amplitudes of this particle are used to compute the aforementioned quantities in the quantum field theory, which for unbounded manifolds can be computed using traditional quantum mechanical path integrals. Hence, first quantisation path integration can be formally used in this framework to extract information in quantum field theories.

For manifolds without boundaries, this worldline formalism is a well established and computationally efficient tool. Since the foundational works of Z. Bern and D. W. Kosower \cite{BernKosower}, and M. J. Strassler \cite{Strassler}, it has been used for computing scattering amplitudes and effective actions for a variety of quantum field theories \cite{Schubert2001}. Scalar, fermionic and vector fields have been established by Strassler in \cite{Strassler} for flat manifolds, and were later analysed in \cite{BastianelliZirotti,BastianelliCorradiniZirotti,BastianelliBenincasaGiombi} for an arbitrary gravitational background. Higher-spin fields in conformally flat manifolds were addressed in \cite{BastianelliCorradiniLatini, Corradini2010, BastianelliBonezziCorradiniLatini}.

In spite of the many models studied and applications considered, field theories in manifolds with boundaries have been more elusive. In order to construct a worldline formulation in these manifolds, one needs to restrict the path integration domain of the point-particle within its boundaries. Furthermore, this restriction also needs to take into account the specific boundary condition imposed on the field, which translates into imposing boundary conditions on the transition amplitudes of the point-particle. Therefore, the problem of formulating a worldline prescription for field theories in bounded manifolds reduces to the problem of constructing a path integral representation for the transition amplitudes in quantum mechanics.

Scalar fields in manifolds with boundaries have been extensively studied. For instance, Dirichlet boundary conditions on a $(D-1)$-dimensional surface $\Sigma$ can be modelled on the whole $\mathbb{R}^D$ through the coupling $\lambda \delta_{\Sigma}(x)$ to a delta-function with support on $\Sigma$: in the limit of infinite coupling $\lambda \rightarrow \infty$ one reproduces Dirichlet boundary conditions. This approach was introduced in the worldline context in \cite{GiesLangfeldMoyaerts} (for a similar mechanism for Neumann boundary conditions, see \cite{FoscoLombardoMazzitelli}). However, this procedure is generally not suitable for usual perturbative calculations: indeed, such procedures would lead to an expansion in positive powers of $\lambda$ and the limit $\lambda \rightarrow \infty$ usually appears to be ill-defined (for a strong coupling approach involving Padé approximants, see \cite{EdwardsGonzalezDominguezHuetTrejo}). An exception occurs for the free scalar field in flat space, where a resummation that leads to the correct heat-trace expansion for the Dirichlet propagator in the limit of infinite coupling is possible \cite{FranchinoMazzitelli} (for a similar resummation involving the Neumann propagator, see \cite{AhmadiniazFranchinoManzoMazzitelli}). If instead of the free scalar field one considers a potential $V(x)$, then the aforementioned resummation for the free Dirichlet propagator can be used to compute the contribution to the heat-trace at different powers of $V(x)$, even for strong coupling \cite{Franchino}. When the coupling constant is left finite, the delta-function coupling to the scalar field reproduces a semitransparent mirror, which was studied in the context of the worldline formalism in \cite{FranchinoPisani}.

A rather different approach to the problem of a bounded field in the worldline formalism involves the use of image charges. This method was used in \cite{BastianelliCorradiniPisani2007,BastianelliCorradiniPisaniSchubert} for a scalar field with either Dirichlet or Neumann boundary conditions on the $D$-dimensional half-space $M = \mathbb{R}^{D-1} \times \mathbb{R}^+$ limited by an infinite flat hyperplane. A generalization of the Neumann case to include Robin boundary conditions was later introduced in \cite{BastianelliCorradiniPisani2008} for the same manifold. This technique strongly relies on the fact that the boundary is flat, so for other types of geometries an adaptation is necessary. In this context, the case of a scalar field confined to a $D$-dimensional ball $B^D$ (that is, the interior of the hypersphere $S^{D-1}$) was analysed for Dirichlet and Neumann boundary conditions in \cite{CorradiniEdwardsHuetManzoPisani}, where a conformal transformation was performed to map the ball into the half-space. This same analysis was extended to include Robin boundary conditions in \cite{TrabajoDiploma}.

In the present work, a similar procedure in terms of image charges is performed for the case of a spinor field minimally coupled to an Abelian gauge field, confined inside certain smooth $2$-dimensional curved manifolds $M$ under MIT bag boundary conditions.\footnote{In some contexts, these are also known as Berry-Mondragón boundary conditions, or infinite-mass boundary conditions.} The procedure, which singles out in the path integral the contributions of worldlines which reach the boundary from those which lie entirely in the bulk, allows one to determine both diagonal and off-diagonal heat-kernel elements. The first ones correspond to closed worldlines within $M$ that allow for the computation of the heat-trace asymptotics ---and, thus, functional determinants---, while the second ones correspond to open worldlines that are useful for the computation of tree-level $n$-point functions within the context of the worldline formalism. Besides constructing a representation for the heat-kernel useful for either closed or open paths, the former scenario is considered in this article for the computation of the first heat-trace asymptotic coefficients (Seeley-DeWitt coefficients).

The reason for considering MIT bag boundary conditions is twofold. On the practical side, these local conditions imply that half of the spinor components satisfy a Dirichlet boundary condition while the other half satisfy a specific Robin boundary condition, and both types of conditions where previously analysed in the context of the worldline formalism for scalar fields. Other local conditions for fermions do not lead to a Robin boundary condition but instead to a more complicated first-order condition, such as the chiral bag boundary condition for which half of the spinor components satisfy a condition involving tangential derivatives, which have not so far been applied within a worldline framework. On the other hand, MIT bag boundary conditions are used in a vast number of physical situations: since they where introduced in \cite{ChodosJaffeJohnsonThornWeisskopf} as part of an extended model for hadrons, they have also been applied to other contexts such as gauged supergravity \cite{BreitenlohnerFreedman}, fermionic billiards \cite{BerryMondragon}, superstrings \cite{SheikhJabbari}, spinning membranes \cite{LuckockMoss} and graphene devices \cite{BeneventanoSantangelo}. It is expected that an analysis of MIT bag boundary conditions in terms of the worldline formalism will allow the implementation of worldline tools in these areas.

The procedure carried in the present article begins by representing the manifold $M$ in coordinates that span the entire half-plane $\mathbb{R} \times \mathbb{R}^+$. In this coordinate representation, the manifold presents a certain metric $g_{\mu \nu}$, and the boundary $\partial M$ is represented via the straight line $\mathbb{R}$. The next step is to duplicate $M$ to build up another region $\widetilde{M} \approx \mathbb{R}^2$ by reflecting the half-plane through its boundary and endowing the resulting full plane with the symmetric extension of the original metric. The curvature of the region $\widetilde{M}$ is different from the curvature of $M$ because the symmetric extension introduces a Heaviside-function on the metric, which is thus non-smooth at the interface $\mathbb{R}$. Besides, path integration of a point-like particle on curved space corresponds to a $0+1$ sigma model which requires certain counterterms ---specific to each regularization--- that are necessary to maintain general coordinate invariance \cite{GreenBook}. In particular, the counterterm required by time-slicing renormalization contains a term proportional to the curvature of $\widetilde{M}$, which is given by a delta-function with support at the interface. As a consequence, the computation of the heat-trace in these coordinates amounts to obtaining the point-particle expectation values of combinations of delta- and Heaviside-functions. Besides these curvature considerations, a symmetric extension of the tangential component of the gauge field in $\widetilde{M}$ is performed, as well as an antisymmetric extension of its normal component. The latter component is made continuous through a gauge fixing condition.

Image charges are used to separate ``direct'' and ``indirect'' contributions to the transition amplitude, according to whether the end-point of the trajectory lies in the physical region $M \subset \widetilde{M}$ or not. Finally, to illustrate the whole procedure, the leading direct and indirect contributions to the heat-trace are computed ---which correspond to the volumes of $M$ and its boundary $\partial M$--- as well as the next-to-leading contribution to obtain the Seeley-DeWitt coefficient $a_2$ which gives the trace anomaly. A worldline formulation in phase space is employed for this application.

The organisation of the article is as follows. In Section \ref{section:effectiveaction} the relation between the heat-trace and the effective action for a quantum $\frac{1}{2}$-spin field coupled to an electromagnetic background is presented. The manifold $M$ and the construction of the doubled manifold $\widetilde{M}$, along with some geometrical properties, is described in Section \ref{SectionGeometry}. After this, geometrical and electromagnetic quantities in $M$ must be ``reflected'' appropriately to $\widetilde{M}$, which is done in Section \ref{ReflectionSection}. This completes the setup in terms of image quantities, from which path integration can be constructed in $\widetilde{M}$. Since the field is fermionic, this path integration must take the Dirac gamma matrices into account. For this purpose, coherent states are chosen in the present article, so they are reviewed in Section \ref{section:CoherentStates}. Next, the transition amplitudes for a fermionic point-particle in curved manifolds are introduced in Section \ref{TransitionAmplitudesSection}, using path integrals in a phase space representation. This completes the setup for using first-quantisation path integrals to compute quantities in a fermionic Quantum Field Theory in manifolds with boundaries. This whole structure is then used considering a specific boundary condition: the MIT bag ones. For completeness, they are described briefly in Section \ref{section:MITbagBC}. Then Section \ref{MainSection} contains the main result of this article, where an ansatz for a fermionic heat-kernel in $M$ under MIT bag boundary conditions is presented. As an application of this construction, the first three Seeley-DeWitt coefficients of the heat-trace are computed in Section \ref{section:SDW}, showing perfect consistency with previous results. Finally, some considerations on the applications of this construction are drawn in Section \ref{section:conclusion}, with particular emphasis on extensions to other boundary conditions for fermions and other kinds of fields, and also the implementations of the procedure to worldline numerical computations.

\section{Effective action}
\label{section:effectiveaction}

Consider a spin$-\frac{1}{2}$ field $\Psi$ of mass $m$ confined to a $D-$dimensional Euclidean manifold $x \in M$, minimally coupled to gravity and to an Abelian gauge field. The action reads

\begin{equation}
    S[A_\mu] = \int_{M} d^Dx \, \sqrt{g} \, \Psi^\dagger \big( -i\slashed{D}-im \big) \Psi \; ,
\end{equation}

\noindent where $g$ is the determinant of the metric in $M$ and\footnote{The convention established for this article is that Greek letters from the middle of the alphabet (such as $\mu$, $\nu$, $\rho$) denote curved indices in the bulk, while Latin letters from the middle of the alphabet (such as $i$, $j$, $k$) denote flat indices in the bulk. Also, Greek letters from the beginning of the alphabet (such as $\alpha$, $\beta$, $\gamma$) denote curved indices in the boundary, while Latin letters from the beginning of the alphabet (such as $a$, $b$, $c$) denote flat indices in the boundary. For a $D-$dimensional manifold with boundaries, bulk indices range from $1$ to $D$ and boundary indices range from $1$ to $D-1$.}.

\begin{equation}\label{GeneralDiracOperator}
    -i\slashed{D} = -i \, e^\mu_k \, \gamma^k \, D_\mu \; , \qquad D_\mu = \partial_\mu + \frac{1}{4} \omega_{\mu ij} \gamma^i \, \gamma^j + i A_\mu \; ,
\end{equation}

\noindent is the (massless) Dirac operator. It contains an Abelian gauge connection $A_\mu$ as well as a spin connection

\begin{equation}\label{VielbeinPostulate}
\omega_{\mu ij} = -\delta_{ik}e^\nu_j \big[ (\partial_\mu e^k_\nu) - \Gamma^\rho_{\mu \nu}e^k_\rho \big] \; ,
\end{equation}

\noindent where $e^\mu_i$ and $e^i_\mu$ are vielbeins such that $g_{\mu \nu} = e^i_\mu e^j_\nu \delta_{ij}$, $\delta^{ij}=e^i_\mu e^j_\nu g^{\mu \nu}$ and $e^\mu_i = g^{\mu \nu} \delta_{ij} e_\nu^j$. The (flat) Dirac matrices $\gamma^i$ are constant and obey the Clifford algebra $\{ \gamma^i , \gamma^j \} = 2\delta^{ij}$. In the present work an even-dimensional Euclidean manifold is considered, so Dirac matrices satisfy $(\gamma^i)^\dagger = \gamma^i$ and the chiral Dirac matrix $\gamma^{\text{ch}} = (-1)^{D/2}\, i \gamma^1\dots\gamma^D$ is well-defined.

The effective action up to one-loop order is

\begin{align}
    \Gamma[A_\mu] & = S[A_\mu] + \hbar \log \text{Det} \big( -i\slashed{D} - im \big) \\
    & = S[A_\mu] + \frac{\hbar}{2} \log \text{Det} \big( -\slashed{D}^2 + m^2 \big)\label{FromFirstToSecondOrderDeterminant} \; .
\end{align}

\noindent Using Schwinger's proper time regularization \cite{Schwinger}, the one-loop correction becomes

\begin{equation}\label{OneLoopEffectiveAction}
    \frac{1}{2} \log \text{Det} \big( -\slashed{D}^2 + m^2 \big) = - \frac{1}{2} \int_0^\infty \frac{dT}{T} \, e^{-Tm^2} \, \text{Tr} \, e^{-T(-\slashed{D}^2)} \; ,
\end{equation}

\noindent which represents the divergent functional determinant in terms of the heat-trace of $-\slashed{D}^2$.

By means of the Schrödinger-Lichnerowicz formula \cite{Schrodinger,Lichnerowicz}, the Dirac operator squared can be written as a coordinate operator in the form

\begin{equation}\label{SchrodingerLichnerowiczFormula}
    -\slashed{D}^2 = -\frac{1}{\sqrt{g}} D_\mu \big(\sqrt{g} g^{\mu \nu} D_\nu \big) - \frac{i}{4} [\gamma^i,\gamma^j] e^\mu_i e^\nu_j F_{\mu \nu} + \frac{1}{4}R \; ,
\end{equation}

\noindent where $R$ is the scalar curvature of the manifold and $F_{\mu \nu} = \partial_\mu A_\nu - \partial_\nu A_\mu$ is the electromagnetic field tensor. In this form it is clear that $-\slashed{D}^2$ is a Laplace-type operator with the fully covariant\footnote{Here, ``fully covariant'' means that the Laplacian is written in terms of covariant derivatives with two types of connections: a curvature connection (the spin connection in this case) and also a gauge connection (which for the present article only includes the Abelian gauge potential).} Laplacian

\begin{equation}\label{FullyCovariantLaplacian}
    \Delta = -\frac{1}{\sqrt{g}} D_\mu \big(\sqrt{g} g^{\mu \nu} D_\nu \big) \; .
\end{equation}

\noindent Under quite general conditions, the heat-trace of $-\slashed{D}^2$ admits the following asymptotic expansion:

\begin{equation}\label{HeatTraceAsymptoticExpansion}
    \text{Tr} \, e^{-T(-\slashed{D}^2)} \sim \frac{1}{(4 \pi T)^{D/2}} \sum_{n=0}^\infty a_n(-\slashed{D}^2 , M) \, T^{n/2} \; ,
\end{equation}

\noindent where the Seeley-DeWitt coefficients $a_n(-\slashed{D}^2 , M)$ can be computed in terms of invariants that only contain the gauge field $A_\mu$, geometric elements of $M$ and its boundary $\partial M$, and parameters appearing in the boundary conditions. Expressions \eqref{OneLoopEffectiveAction} and \eqref{HeatTraceAsymptoticExpansion} show that the coefficients $a_n(-\slashed{D}^2 , M)$ with $0 \leq n \leq D$ give the one-loop divergences of the effective action $\Gamma[\Psi,\Psi^\dagger]$, while the coefficients with $n>D$ contribute to the finite part of $\Gamma[\Psi,\Psi^\dagger]$. In particular, the coefficient $a_0(-\slashed{D}^2 , M) \sim \text{Vol}(M)$ is given only by the volume of the manifold $M$ and does not depend on any other geometric property of space. Similarly, $a_1(-\slashed{D}^2 , M) \sim \text{Vol}(\partial M)$ depends only on the volume of the boundary $\partial M$ as well as the boundary condition \cite{HeatKernelManual}.

This is just one example of the applications of heat-kernel techniques to the perturbative study of quantum field theories. In this article it is shown how to compute the coefficients $a_n(-\slashed{D}^2,M)$ for certain smooth 2-dimensional manifolds using worldline techniques.

\section{Geometry of $M$ and the `doubled' manifold $\widetilde{M}$}\label{SectionGeometry}

\begin{figure}[ht!]
    \centering
    \begin{subfigure}{.45\textwidth}
        \centering
        \includegraphics[width=.85\linewidth]{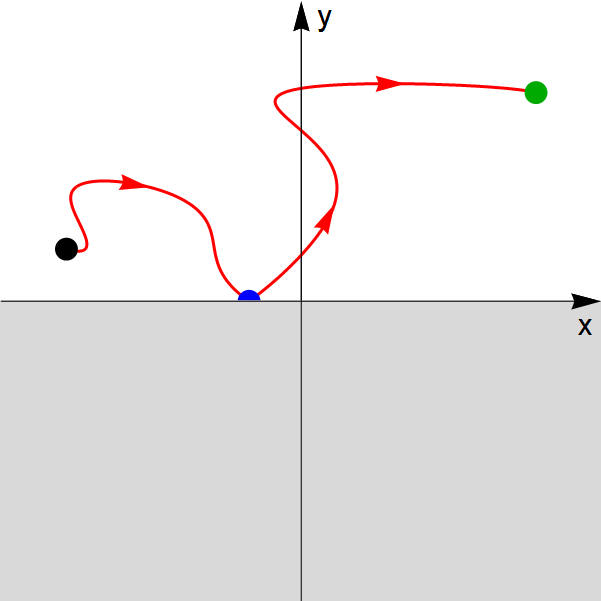}
        \caption{}
        \label{Figure:PathsWithouthReflection}
    \end{subfigure}
    \begin{subfigure}{.45\textwidth}
        \centering
        \includegraphics[width=.85\linewidth]{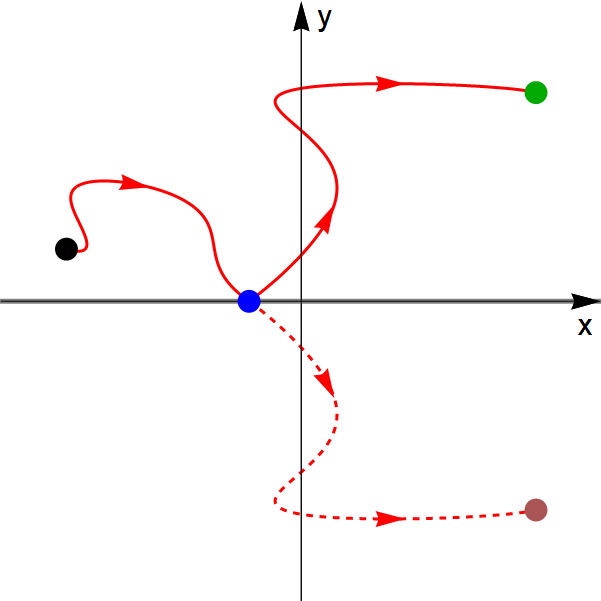}
        \caption{}
        \label{Figure:PathsWithReflection}
    \end{subfigure}
    \caption{Worldlines in two dimensions. Picture \ref{Figure:PathsWithouthReflection} shows a worldline in the manifold $M$ from initial point $\bullet$ (black) to final point \textcolor{Green}{$\bullet$} (green), hitting the boundary once in the intermediate point \textcolor{blue}{$\bullet$} (blue). Picture \ref{Figure:PathsWithReflection} represents $\widetilde{M}$, where the boundary turns into an interface. Point \textcolor{IndianRed}{$\bullet$} (pink) in the lower half-plane is the reflection of \textcolor{Green}{$\bullet$} (green) along this interface. A typical curve from $\bullet$ (black) to \textcolor{Green}{$\bullet$} (green) that hits the boundary once at \textcolor{blue}{$\bullet$} (blue) has a corresponding curve from $\bullet$ (black) to \textcolor{IndianRed}{$\bullet$} (pink): the segment from $\bullet$ (black) to \textcolor{blue}{$\bullet$} (blue) remains the same, while the segment from \textcolor{blue}{$\bullet$} (blue) to \textcolor{Green}{$\bullet$} (green) gets symmetrically reflected from \textcolor{blue}{$\bullet$} (blue) to \textcolor{IndianRed}{$\bullet$} (pink). The contribution of the original curve is called `direct', and the contribution of the second one is called `indirect'. Both curves measure equally.}
    \label{Figure:PathsIn2D}
\end{figure}

In Section \ref{TransitionAmplitudesSection} it is shown that the heat-trace asymptotics \eqref{HeatTraceAsymptoticExpansion} for the Laplace-type operator $-\slashed{D}^2$ in $M$ is determined by the path integral over closed trajectories of a point-particle. In order to study the dynamics of this particle, it is convenient to identify $M$ with a $2$-dimensional half-plane which is then embedded into a whole plane $\mathbb{R}^2$, denoted $\widetilde{M}$, that represents two copies of the original manifold $M$ glued together along the interface $\partial M$ as in Figure \ref{Figure:PathsIn2D}.

Hence, consider a smooth two-dimensional manifold representable in half-plane coordinates $(x,y)$ such that $-\infty < x < \infty$ and $0 \leq y < \infty$. In particular, $y = 0$ represents the boundary $\partial M$. Let $g_{\mu \nu}$ be the metric in this coordinate representation, which will be considered as non-singular and diagonal, so in its matrix form one can write

\begin{equation}\label{MetricMatrix}
    g_{\mu \nu} = \begin{pmatrix}
        h & 0 \\
        0 & g/h
    \end{pmatrix}
\end{equation}

\noindent where both $h \equiv h(x,y)$ and $g \equiv g(x,y)$ are non-singular functions without zeros. Both of them have a straightforward geometrical interpretation: on one hand, when evaluated at $y=0$, the function $h(x,0)$ is the metric of the boundary $\partial M$, so $h(x,y)$ is an extension of $h(x,0)$ to the bulk of $M$. On the other hand, $g = \det \, g_{\mu \nu}$.

Next, extend this metric to the whole plane $\mathbb{R}^2$ by making a symmetric reflection with respect to the line $y=0$. This ``doubled manifold'', which is denoted as $\widetilde{M}$, has metric

\begin{equation}\label{Metric}
    \widetilde{g}_{\mu \nu}(x,y) = g_{\mu \nu}(x,|y|) \; .
\end{equation}

\noindent note that the doubled metric may no longer be analytic for it may have a discontinuous normal derivative at the fixed points $y=0$. The corresponding integration measure is

\begin{equation}
    \sqrt{\widetilde{g}(x,y)} = \sqrt{g(x,|y|)} \; .
\end{equation}

\noindent Denoting objects computed in $\widetilde{M}$ with a tilde `` $\, \widetilde{} \,$ '', one can express the Christoffel symbols as

\begin{equation}
    \widetilde{\Gamma}^\rho_{\mu \nu} = \Gamma^\rho_{\mu \nu} - \theta(-y) \, g^{\rho \sigma} \Big[ \delta^2_\mu \, \partial_2 g_{\nu \sigma} + \delta^2_\nu \, \partial_2 g_{\mu \sigma} - \delta^2_\sigma \, \partial_2 g_{\mu \nu} \Big]
\end{equation}

\noindent where $\theta(y)$ is the Heaviside step function. In this and all subsequent geometrical expressions --unless something different is explicitly stated--, every object computed in $\widetilde{M}$ is evaluated at $(x,y)$ and objects computed in $M$ are evaluated at $(x,|y|)$. The scalar curvature and the full contraction of two Christoffel symbols are

\begin{align}
    \widetilde{R} & = R
    - \frac{2}{g} \delta(y) (\partial_2 h) \; ,\label{ScalarCurvatureFirst}\\
    \widetilde{\Gamma}^2 & \equiv \widetilde{g}^{\mu \nu} \widetilde{\Gamma}^\sigma_{\mu \rho} \widetilde{\Gamma}^\rho_{\nu \sigma} = g^{\mu \nu} \Gamma^\sigma_{\mu \rho} \Gamma^\rho_{\nu \sigma} \equiv \Gamma^2 \label{GammaGammaCounterterm}\; .
\end{align}

\noindent note that the difference between the scalar curvatures of $M$ and $\widetilde{M}$ has support at the boundary.

Path integration in curved manifolds requires the introduction of an additional counterterm potential to ensure coordinate invariance. In the present manuscript, phase space path integration will be used, which in curved spaces without boundaries is suitably described in terms of a Time Slicing formulation which involves the counterterm \cite{GreenBook}

\begin{equation}\label{CountertermsGeneralEquation}
    \Delta H_{TS} = -\frac{1}{4}\widetilde{R} + \frac{1}{4} \widetilde{\Gamma}^2 \; .
\end{equation}

\noindent Time Slicing construction for the computation of transition amplitudes is suitably defined in the manifold $\widetilde{M}$, which has no boundaries. Therefore, the idea of the present article is to use path integrals in $\widetilde{M}$ and relate them to the transition amplitudes in $M$ under MIT bag boundary conditions.

Going back to $M$, one needs to consider the inward-pointing normal unit vector $n^\mu$ at the boundary and extend it in some way to the interior of the manifold. From the metric \eqref{MetricMatrix} it follows

\begin{equation}\label{NormalUnitVectorAtBoundary}
    n_\mu \big|_{\partial M} = \sqrt{g/h}\Big|_{\partial M} \, \delta^2_\mu \; .
\end{equation}

\noindent As for the bulk of the manifold, a parallel transport extension along geodesics normal to the boundary will be considered. Therefore, $n_\mu$ obeys in $M$ the parallel transport equation

\begin{equation}\label{NormalExtensionEquation}
    n^\nu \nabla_\nu n^\mu = 0 \; .
\end{equation}

The second fundamental form in $\partial M$ has a single component $L_{11}$ and is given by

\begin{equation*}
    L_{11} \equiv -\nabla_1 n_1 \big|_{\partial M} = -\sqrt{g/h} \, \Gamma^2_{11} \Big|_{\partial M} = -\frac{1}{2} \sqrt{h/g} \,(\partial_2 h) \Big|_{\partial M} \; .
\end{equation*}

\noindent Then the extrinsic curvature $L$, which is defined as the trace of the second fundamental form and is an invariant in $\partial M$, is

\begin{equation}\label{ExtrinsicCurvature}
L \equiv g^{11}L_{11} = -\frac{1}{2} \frac{1}{\sqrt{h \, g}} \,(\partial_2 h) \Big|_{\partial M} \; .
\end{equation}

\noindent Then \eqref{ScalarCurvatureFirst} becomes

\begin{equation}\label{ScalarCurvatureSecond}
    \widetilde{R} = R + 4 L \sqrt{\frac{h}{g}} \, \delta (y) \; ,
\end{equation}

\noindent which is an expression relating scalar invariants in $M$, $\partial M$ and $\widetilde{M}$.

Going back to $M$, one needs to analyse the vielbeins $e^i_\mu$ that are present in the Dirac operator squared, both explicitly an implicitly through the spin connection $\omega_{\mu i j}$. In the present article, the following choice will be used to set the vielbeins: at any point in $M$, let $e^{i=2}_\mu$ be equal to the normal unit vector $n_\mu$. It follows that $e^{i=1}_\mu$ must be equal to a unit vector perpendicular to $n_\mu$. This leads to a sign (directional) ambiguity that is solved by picking the direction that obeys the following 2-dimensional identity:

\begin{equation}\label{VielbeinsAndMetricDeterminant}
    \text{det } e^i_\mu = \sqrt{g} \; .
\end{equation}

\noindent At this point it must be remarked that with the aforementioned vielbein choice, every geometrical element in \eqref{SchrodingerLichnerowiczFormula} must be non-singular. This ensures that the transition amplitudes to be constructed are well-defined in the entire manifold $M$.

Finally, to simplify expressions in two dimensions one could use the antisymmetry of the spin connection with respect to the flat indices to define $\omega_{\mu} \equiv \omega_{\mu 12}$ (`$1$' and `$2$' are flat indices in this definition), which from \eqref{VielbeinPostulate} can be written as $\omega_\mu = -e^\nu_{i=2} \nabla_\mu e^{i=1}_\nu = e^{i=1}_\nu \nabla_\mu e^\nu_{i=2}$. This definition implies $n^\mu \omega_\mu = 0$, as one can easily see using \eqref{NormalExtensionEquation} and the vielbein choice discussed in the previous paragraph. In particular, this means that $\omega_2 |_{y=0} = 0$.

\section{Symmetry of functions and operators reflected to the lower half-plane}\label{ReflectionSection}

In the spirit of the method of images, one expects paths in the upper half-plane to have the same measure as reflected paths in the lower half-plane. To ensure this, several considerations about how to extend functions and operators must be taken into account. This task is carried out in the present Section. Firstly, the extension of geometrical quantities in $M$ such as the metric $g_{\mu \nu}$, the normal vector $n^\mu$, the Riemann covariant derivative $\nabla_\mu$ and the spin connection $\omega_{\mu i j}$ is performed. Secondly, electromagnetic quantities such as the background gauge field $A_\mu$ and the field tensor $F_{\mu \nu}$ are taken in consideration by analysing the geometry of gauge invariance. The results obtained lead to a natural extension of operators that depend on this geometrical and electromagnetic factors.

\subsection{Reflection of geometrical quantities}\label{ReflectionGeometrySection}

If paths in the upper half-plane and their reflected paths in the lower half-plane must have the same value, then the line element $ds = g_{\mu \nu} x^\mu x^\nu$ must be the same in both half-planes. For this purpose, the metric $g_{\mu \nu}$ was extended symmetrically. In mathematical form and using \eqref{MetricMatrix}, this reads

\begin{equation}
\begin{aligned}
    g(x,y) & = g(x,-y) \; ,\\
    h(x,y) & = h(x,-y) \; ,
\end{aligned}
\end{equation}

\noindent where this extension means both functions have domain in $\widetilde{M}$.\footnote{Unless otherwise stated, the domain of every function $f$ in the present Section is $\widetilde{M}$, and is an extension of the corresponding function $f$ defined in $M$. They should not be confused with the function $\widetilde{f}$, whose domain is also $\widetilde{M}$. For example, the scalar curvature $R$ with domain in $\widetilde{M}$ is obtained by first computing $R$ in $M$ (that is, computing the corresponding derivatives of $g_{\mu \nu}$ in $M$) and then extending the result symmetrically to $\widetilde{M}$, whereas $\widetilde{R}$ is obtained by first extending the metric $g_{\mu \nu}$ symmetrically to $\widetilde{M}$ and then computing the corresponding derivatives.} To ensure that the lower half-plane is a physically equivalent reflection of the upper half-plane, every spinor in the theory is chosen to be reflected symmetrically as well:

\begin{equation}\label{ReflectionSpinor}
\begin{aligned}
    \Psi(x,y) & = \Psi(x,-y) \; ,\\
    \Psi^\dagger(x,y) & = \Psi^\dagger(x,-y) \; .
\end{aligned}
\end{equation}

Next, consider the partial derivative operator $\partial_\mu$ acting on spinors. Since the spinors are reflected symmetrically, their partial derivative in the $y$-direction should be antisymmetrical:

\begin{equation}\label{ReflectionPartialDerivative}
\begin{aligned}
    \partial_1 |_{y>0} & = \partial_1 |_{y<0} \; ,\\
    \partial_2 |_{y>0} & = -\partial_2 |_{y<0} \; .
\end{aligned}
\end{equation}

\noindent From metric \eqref{MetricMatrix} one can see that if an even (odd) number of indices in the Christoffel symbol are equal to 2, then the Christoffel symbol is proportional to a derivative in the $x$-direction ($y$-direction) of a metric component. Then, since the metric is reflected symmetrically, one gets

\begin{equation}\label{ReflectionChristoffel}
\begin{aligned}
    \Gamma^\rho_{\mu \nu}(x,y) = \Gamma^\rho_{\mu \nu}(x,-y) & \qquad \text{if an even number of indices are equal to 2} \; ,\\
    \Gamma^\rho_{\mu \nu}(x,y) = -\Gamma^\rho_{\mu \nu}(x,-y) & \qquad \text{if an odd number of indices are equal to 2} \; ,
\end{aligned}
\end{equation}

\noindent which leads to

\begin{equation}\label{ReflectionRiemannDerivative}
\begin{aligned}
    \nabla_1 |_{y>0} & = \nabla_1 |_{y<0} \; ,\\
    \nabla_2 |_{y>0} & = -\nabla_2 |_{y<0} \; .
\end{aligned}
\end{equation}

\noindent Reflections \eqref{ReflectionChristoffel} also imply that the scalar curvature $R$ and $\Gamma^2$ are reflected symmetrically.

Moving on to the normal unit vector $n^\mu$, the chosen extension consist in symmetrizing the component $n^2$ and antisymmetrizing the component $n^1$, which has the advantage that the normal vector is not discontinuous at the interface $y=0$. In mathematical form,

\begin{equation}
\begin{aligned}
    n^1(x,y) & = -n^1(x,-y) \; ,\\
    n^2(x,y) & = n^2(x,-y) \; .
\end{aligned}
\end{equation}

\noindent Since $n_\mu = e^{i=2}_\mu$ in the upper half-plane, the same vielbein choice can be set for the lower half-plane, and then

\begin{equation}\label{ReflectionVielbein2}
\begin{aligned}
    e^{i=2}_{\mu=1}(x,y) & = -e^{i=2}_{\mu=1}(x,-y) \; ,\\
    e^{i=2}_{\mu=2}(x,y) & = e^{i=2}_{\mu=2}(x,-y) \; .
\end{aligned}
\end{equation}

\noindent Retaining \eqref{VielbeinsAndMetricDeterminant} in the lower half-plane leads to the reflection of $e^{i=1}_\mu$ according to

\begin{equation}\label{ReflectionVielbein1}
\begin{aligned}
    e^{i=1}_{\mu=1}(x,y) & = e^{i=1}_{\mu=1}(x,-y) \; ,\\
    e^{i=1}_{\mu=2}(x,y) & = -e^{i=1}_{\mu=2}(x,-y) \; .
\end{aligned}
\end{equation}

\noindent Recalling the expression $\omega_{\mu} = -e^\nu_{i=2} \nabla_\mu e^{i=1}_\nu$ and using \eqref{ReflectionRiemannDerivative}, \eqref{ReflectionVielbein2} and \eqref{ReflectionVielbein1} it follows

\begin{equation}\label{ReflectionSpinConnection}
    \begin{aligned}
        \omega_1(x,y) & = -\omega_1(x,-y) \; ,\\
        \omega_2(x,y) & = \omega_2(x,-y) \; .
    \end{aligned}
\end{equation}

\subsection{Reflection of electromagnetic quantities}\label{ReflectionElectroSection}

Consider first the local Abelian gauge transformation of the spinor field

\begin{equation}\label{LocalGaugeTransformation}
    \Psi(x,y) \rightarrow e^{i \alpha(x,y)} \Psi(x,y) \; .
\end{equation}

\noindent If the symmetry property \eqref{ReflectionSpinor} is meant to be satisfied for an arbitrary gauge, it follows that

\begin{equation}
    \alpha(x,y) = \alpha(x,-y) \; .
\end{equation}

Now let $v^\mu$ be an arbitrary unit vector. Since \eqref{LocalGaugeTransformation} is a local transformation, $\Psi(x,y)$ and $\Psi(x + \varepsilon v^1,y + \varepsilon v^2)$ transform differently for $\varepsilon \neq 0$. Therefore, a partial derivative is not a very sensible way to compare both quantities because it depends on the difference $\Psi(x + \varepsilon v^1,y + \varepsilon v^2) - \Psi(x,y)$. To construct a gauge covariant derivative $\mathcal{D}_\mu$, one then introduces a comparator $U(x+\varepsilon v^1,y+\varepsilon v^2;x,y)$ that transforms under

\begin{equation}
    U(x+\varepsilon v^1,y+\varepsilon v^2;x,y) \rightarrow e^{i\alpha(x+\varepsilon v^1,y+\varepsilon v^2)} U(x+\varepsilon v^1,y+\varepsilon v^2;x,y) e^{-i\alpha(x,y)} \; ,
\end{equation}

\noindent so both $\Psi(x + \varepsilon v^1,y + \varepsilon v^2)$ and $U(x+\varepsilon v^1,y+\varepsilon v^2;x,y) \Psi(x,y)$ obey the same transformation law. At zero separation $\varepsilon = 0$, one sets $U(x,y;x,y) = 1$. Then one defines the gauge covariant derivative in the direction of $v^\mu$ as

\begin{equation}
    v^\mu \mathcal{D}_\mu \Psi(x,y) = \lim_{\varepsilon \rightarrow 0} \frac{1}{\varepsilon} \Big[ \Psi(x + \varepsilon v^1,y + \varepsilon v^2) - U(x+\varepsilon v^1,y+\varepsilon v^2;x,y) \Psi(x,y) \Big] \; .
\end{equation}

Considering $U(x+\varepsilon v^1,y+\varepsilon v^2;x,y) \Psi(x,y)$ as a spinor, it must satisfy the reflection condition \eqref{ReflectionSpinor}. Then

\begin{equation}\label{ReflectionComparator}
    U(x+\varepsilon v^1,y+\varepsilon v^2;x,y) = U(x+\varepsilon v^1,-y-\varepsilon v^2;x,-y) \; .
\end{equation}

\noindent Therefore

\begin{equation}
    v^\mu \mathcal{D}_\mu \Psi(x,-y) = \Tilde{v}^\mu \mathcal{D}_\mu \Psi(x,y)
\end{equation}

\noindent where $(\Tilde{v}^1,\Tilde{v}^2) = (v^1,-v^2)$. In particular, setting $v^\mu \propto \delta^\mu_1$ or $v^\mu \propto \delta^\mu_2$ one gets

\begin{equation}\label{ReflectionGaugeDerivative}
\begin{aligned}
    \mathcal{D}_1 |_{y>0} & = \mathcal{D}_1 |_{y<0} \; ,\\
    \mathcal{D}_2 |_{y>0} & = -\mathcal{D}_2 |_{y<0} \; .
\end{aligned}
\end{equation}

One can now expand

\begin{equation}\label{ComparatorWithGaugeField}
    U(x+\varepsilon v^1,y+\varepsilon v^2;x,y) = 1 - i \varepsilon v^\mu A_\mu(x,y) + \mathcal{O}(\varepsilon^2) \; ,
\end{equation}

\noindent where $A_\mu$ is the electromagnetic gauge field, to obtain

\begin{equation}\label{GaugeCovariantDerivative}
    \mathcal{D}_\mu = \partial_\mu + i A_\mu \; .
\end{equation}

\noindent From \eqref{ReflectionComparator} and \eqref{ComparatorWithGaugeField}, or from \eqref{GaugeCovariantDerivative}, \eqref{ReflectionGaugeDerivative} and \eqref{ReflectionPartialDerivative}, one finds

\begin{equation}\label{ReflectionGaugeField}
    \begin{aligned}
        A_1(x,y) & = A_1(x,-y) \; ,\\
        A_2(x,y) & = -A_2(x,-y) \; .
    \end{aligned}
\end{equation}

In a theory where the partial derivative $\partial_2 A_2$ is found, such as in the Laplacian \eqref{FullyCovariantLaplacian}, it is convenient to chose a gauge that ensures the continuity of $A_2$ at $y=0$ to avoid the unnecessary appearance of some delta functions $\delta(y)$. Therefore, the gauge `semi-fixing'\footnote{The term `semi-fixing' comes from the fact that \eqref{GaugeSemiFixing} is not a condition imposed in the entire manifold $\widetilde{M}$ but only at the interface $y=0$.} condition

\begin{equation}\label{GaugeSemiFixing}
    A_2(x,0) = 0
\end{equation}

\noindent will be imposed throughout this article.

Finally, the electromagnetic tensor field is defined as

\begin{equation}
    F_{\mu \nu} \Psi(x,y) = -i[\mathcal{D}_\mu , \mathcal{D}_\nu] \Psi(x,y) \; .
\end{equation}

\noindent Since it is antisymmetric in its indices, the only independent component in two dimensions is $F \equiv F_{12}$. Using \eqref{ReflectionGaugeDerivative} it follows

\begin{equation}\label{ReflectionElectromagneticField}
    F(x,y) = -F(x,-y) \; .
\end{equation}

\subsection{Reflection of operators}\label{ReflectionOperatorsSection}

Since the goal of the present manuscript is to express transition amplitudes in the manifold $M$ (with boundary) in terms of transition amplitudes in the manifold $\widetilde{M}$ (without boundaries), then every operator $\hat{O} : \mathcal{H}(M) \rightarrow \mathcal{H}(M)$ defined in the Hilbert space $\mathcal{H}(M)$ must be somehow extended to a new operator $\hat{O} : \mathcal{H}(\widetilde{M}) \rightarrow \mathcal{H}(\widetilde{M})$ defined in the Hilbert space $\mathcal{H}(\widetilde{M})$.\footnote{To simplify the notation in the rest of the article, the name of the operator $\hat{O}$ has not been changed after being extended from $\mathcal{H}(M)$ to $\mathcal{H}(\widetilde{M})$.} This extension is written as

\begin{equation}\label{ReflectionOperators}
    \hat{O} = \begin{cases}
        \hat{O}^{>} & \text{if } y>0 \\
        \hat{O}^{<} & \text{if } y<0
    \end{cases} \; ,
\end{equation}

\noindent where $\hat{O}^>$ is the same operator as $\hat{O} : \mathcal{H}(M) \rightarrow \mathcal{H}(M)$, and $\hat{O}^<$ is computed by copying $\hat{O}^>$ to the lower half-plane and introducing the reflection properties of Sections \ref{ReflectionGeometrySection} and \ref{ReflectionElectroSection}. This will ensure that any transition amplitude contained within the upper half-plane measure the same as the reflected transition amplitude contained within the lower half-plane.

To set ideas, consider the Dirac operator squared \eqref{SchrodingerLichnerowiczFormula}. Using \eqref{MetricMatrix} one can write, in two dimensions,

\begin{align}
    -\slashed{D}^2 & = -\frac{1}{\sqrt{g}} D_\mu \big(\sqrt{g} g^{\mu \nu} D_\nu \big) + g^{-1/2} F \, \gamma^{\text{ch}} + \frac{1}{4}R \; ,\\
    D_\mu & = \partial_\mu + \frac{i}{2} \omega_\mu \gamma^\text{ch} + i A_\mu \; .
\end{align}

\noindent Of course, this is defined in $\mathcal{H}(M)$. The extension to $\mathcal{H}(\widetilde{M})$ reads

\begin{align}
    -\slashed{D}^2 & = \begin{cases}
        (-\slashed{D}^2)^{>} & \text{if } y>0 \\
        (-\slashed{D}^2)^{<} & \text{if } y<0
    \end{cases} \; ,\label{DiracSquaredReflected}\\
    (-\slashed{D}^2)^{>} & = -\frac{1}{\sqrt{g}} D_\mu \big(\sqrt{g} g^{\mu \nu} D_\nu \big) + g^{-1/2} F \, \gamma^{\text{ch}} + \frac{1}{4}R \; ,\label{DiracSquaredAtUpperHalfPlane}\\
    D_\mu & = \partial_\mu + \frac{i}{2} \omega_\mu \gamma^\text{ch} + i A_\mu \; ,\label{DiracSquaredAtUpperHalfPlaneBis}\\
    (-\slashed{D}^2)^{<} & = -\frac{1}{\sqrt{g}} \overline{D}_\mu \big(\sqrt{g} g^{\mu \nu} \overline{D}_\nu \big) - g^{-1/2} F \, \gamma^{\text{ch}} + \frac{1}{4}R \; ,\label{DiracSquaredAtLowerHalfPlane}\\
    \overline{D}_\mu & = \partial_\mu - \frac{i}{2} \omega_\mu \gamma^\text{ch} + i A_\mu \; .\label{DiracSquaredAtLowerHalfPlaneBis}
\end{align}

\noindent That is, $(-\slashed{D}^2)^{<}$ is operationally similar to $(-\slashed{D}^2)^{>}$, which corresponds to the Dirac operator squared in the upper half-plane, except for the change of sign in the electromagnetic field tensor $F$ and the spin connection $\omega_\mu$.

\section{Fermionic coherent states}\label{section:CoherentStates}

If one aims to consider transition amplitudes of a point-particle on the manifold $M$, it is necessary to integrate not only on configuration space $|x \rangle$ or phase space $|x \rangle \otimes |p \rangle$, but also on the internal degrees of freedom. For fermions, these degrees of freedom are given by the different spinor components. Two approaches exist for this purpose, namely the ``Weyl symbol'' method \cite{Fradkin,BerezinMarinov,FradkinGitman,HenneauxTeitelboim,AhmadiniazBandaGuzmanBastianelliCorradiniEdwardsSchubert} and the ``coherent state'' method \cite{OhnukiKashiwa,BordiCasalbuoni,HentyHoweTownsend,vanHolten,DHokerGagne1,DHokerGagne2,Bhattacharya}. The latter will be used in this manuscript and goes as follows.\footnote{The present description of the coherent state method follows \cite{GreenBook}.}

Firstly, recall the extension \eqref{DiracSquaredReflected} of the Dirac operator squared, written entirely in terms of the fermionic operators $1$ and $\gamma^\text{ch}$. Instead of representing operators acting on a spinor $\Psi$ through these matrices, define the operators

\begin{equation}
    \psi \equiv \frac{1}{2}(\gamma^1 + i \gamma^2) \; , \qquad \psi^\dagger \equiv \frac{1}{2}(\gamma^1 - i \gamma^2) \; ,
\end{equation}

\noindent which obey the anticommutation relations $\{ \psi, \psi^\dagger \} = 1$ and $\{ \psi, \psi \} = \{ \psi^\dagger, \psi^\dagger \} = 0$. Then, instead of describing the fermionic parts of $\Psi$ through the spinor components $\Psi_1$ and $\Psi_2$, one can construct the ``vacuum state'' $|0 \rangle$ and the ``excited state'' $|1 \rangle = \psi^\dagger \, |0 \rangle$ such that the vacuum obeys $\psi \, |0 \rangle = 0$ and $\langle 0 | 0 \rangle = 1$. One can now define

\begin{equation}\label{CoherentStates}
    \begin{aligned}
        |\eta\rangle & \equiv e^{-\eta \psi^\dagger} |0\rangle \equiv (1 - \eta \psi^\dagger)|0\rangle \equiv |0\rangle - \eta |1\rangle \; , \\
        \langle \Bar{\eta} | & \equiv \langle 0 | \, e^{-\psi \Bar{\eta}} \equiv \langle 0 | (1 - \psi \Bar{\eta}) \equiv \langle 0 | - \langle 1 | \Bar{\eta} \; .
    \end{aligned}
\end{equation}

\noindent Both $\eta$ and $\Bar{\eta}$ are Grassmann numbers defined in such a way as to commute with the vacuum state $|0\rangle$ and to anticommute with both operators $\psi$ and $\psi^\dagger$. These states are coherent with respect to the operators $\psi$ and $\psi^\dagger$:

\begin{equation}
    \begin{aligned}
        \psi |\eta \rangle & = \eta |\eta \rangle = |\eta \rangle \eta \; , \\
        \langle \Bar{\eta} | \psi^\dagger & = \langle \Bar{\eta} | \Bar{\eta} = \Bar{\eta} \langle \Bar{\eta} | \; .
    \end{aligned}
\end{equation}

\noindent From these definitions, it can be proved that the trace of any zero-degree operator $\hat{O} \equiv O (\psi,\psi^\dagger)$ is given by

\begin{equation}
    \text{Tr}(\hat{O}) = \int d\eta \, d\bar\eta \, e^{\bar\eta \eta} \langle \bar\eta | \hat{O} | \eta \rangle = \langle 0 | \hat{O} | 0 \rangle + \langle 1 | \hat{O} | 1 \rangle \; ,
\end{equation}

\noindent where Grassmann integration obeys

\begin{equation}
    \begin{aligned}
        \int d\eta \, 1 = \int d\Bar{\eta} \, 1 = 0 \; , \qquad \int d\eta \, \eta = \int d\Bar{\eta} \, \Bar{\eta} = 1 \; .
    \end{aligned}
\end{equation}

\noindent Finally, the chiral Dirac matrix is identified with the operator

\begin{equation}
    \gamma^\text{ch} = -i\gamma^1 \gamma^2 \equiv \psi \psi^\dagger - \psi^\dagger \psi \equiv 2(\psi \psi^\dagger)_A \; ,
\end{equation}

\noindent where the subscript $A$ stands for ``antisymmetrized''.

\section{Transition amplitudes of a fermionic point-particle}\label{TransitionAmplitudesSection}

The transition amplitudes of a fermionic point-particle in a two-dimensional manifold are now at hand. Let $H$ be the Hamiltonian of the particle. Now symmetrize it in terms of the bosonic operators $x$ and $p$ and antisymmetrize it in terms of the fermionic operators $\psi$ and $\psi^\dagger$. This version of $H$, which is written as $H_W = [g^{\mu \nu} p_\mu p_\nu]_S + V_W$, is the ``Weyl ordered'' Hamiltonian of the particle, where $V_W$ is the corresponding effective potential, also Weyl ordered.\footnote{An arbitrary potential will be considered in the present Section.} The subscript $S$ stands for ``symmetrized''.

Consider the evolution of this particle at time $T$ from state $|x,y,\eta \rangle$, with fermionic coherent number $\eta$ and located at position $(x,y)$, to state $\langle x',y',\bar\eta |$. For convenience, the corresponding trajectories will be described as $x(t) = x_0(t) + q^1(t)$ and $y(t) = y_0(t) + q^2(t)$, where $x_0(t)$ and $y_0(t)$ consist on the straight line

\begin{equation}
    x_0(t) = x + \frac{t}{T}(x'-x) \quad \text{and} \quad y_0(t) = y + \frac{t}{T}(y'-y)
\end{equation}

\noindent that connects the point $\big(x_0(0) = x \, , \, y_0(0) = y \big)$ with $\big(x_0(T) = x' \, , \, y_0(T) = y' \big)$, and $\big( q^1(t) , q^2(t) \big)$ representing quantum fluctuations under homogeneous Dirichlet conditions $q^1(0) = q^2(0) = q^1(T) = q^2(T) = 0$. Similarly, paths in the coherent space will be described as $\eta(t) = \eta + \psi(t)$ and $\bar\eta(t) = \bar\eta + \psi^\dagger(t)$, where the quantum fluctuations obey $\psi(0) = \psi^\dagger(T) = 0$. Then the transition amplitudes can be represented in terms of a bosonic path integral in phase space and a fermionic path integral in coherent-space \cite{GreenBook}:

\begin{multline}
    \langle x',y',\bar\eta \, | \, e^{-TH} \, | x,y,\eta \rangle = \Big( \sqrt{g(x,y)} \sqrt{g(x',y')} \Big)^{-1/2} e^{\bar\eta \eta} \times \\
    \int \mathcal{D}p \mathcal{D}q \mathcal{D}\psi^\dagger \mathcal{D}\psi \; e^{-\int_0^T dt \big\{ g^{\mu \nu} p_{\mu} p_{\nu} - i p_1 \big( \frac{x'-x}{T} + \Dot{q}^1 \big) - i p_2 \big( \frac{y'-y}{T} + \Dot{q}^2 \big) + \psi^\dagger \Dot{\psi} + V_W \big\} } \; .
\end{multline}

\noindent In the integrand, both $g^{\mu \nu}$ and $V_W$ are evaluated at $x_0(t) + q^1(t)$ and $y_0(t) + q^2(t)$. The effective potential $V_W$ is also evaluated at $\eta + \psi(t)$ and $\bar\eta + \psi^\dagger(t)$. Evaluating the inverse metric at a fixed point $(x,y)$, which in the present manuscript will be taken as the initial point, path integrals turn out to be normalised according to

\begin{align}
    \int \mathcal{D}p\mathcal{D}q \, e^{-\int_0^T dt \big\{ g^{\mu \nu}(x,y) \, p_{\mu} p_{\nu} - i p_\mu \Dot{q}^\mu \big\} } & = \frac{\sqrt{g(x,y)}}{4 \pi T} \; ,\\
    \int \mathcal{D}\psi^\dagger \mathcal{D}\psi \, e^{-\int_0^T dt \, \psi^\dagger \Dot{\psi}} & = 1 \; .
\end{align}

In order to keep track of the different powers of the (small) variable $T$, it is useful to turn to dimensionless quantities: $t \rightarrow Tt$, $p_\mu(t) \rightarrow p_\mu(t)/\sqrt{T}$ and $q^\mu(t) \rightarrow \sqrt{T} q^\mu(t)$. Next, it is convenient to expand $g^{\mu \nu}$ in the kinetic factor around the initial point $(x,y)$ as

\begin{equation}\label{DeltaCurvature}
    \begin{aligned}
    g^{\mu \nu} & \Big( x_0(t) + \sqrt{T} q^1(t) \, , \, y_0(t) + \sqrt{T} q^2(t) \Big) = g^{\mu \nu}(x,y) + \Delta g^{\mu \nu} \; ,\\
    \Delta g^{\mu \nu} & \equiv g^{\mu \nu} \Big( x + t(x'-x) + \sqrt{T} q^1(t) \, , \, y + t(y'-y) + \sqrt{T} q^2(t) \Big) - g^{\mu \nu}(x,y) \; ,
    \end{aligned}
\end{equation}

\noindent where $\Delta g^{\mu \nu}$ is meant to be expanded in a power series around $(x,y)$. It is also convenient to shift the momentum variables in the exponent and complete squares to get rid of the factors $p_1 \, (x'-x)$ and $p_2 \, (y-y')$. Defining

\begin{equation*}
    \Delta x^\mu = (x'-x,y'-y) \; , \quad
    \xi^2 = g_{\mu \nu}(x,y) \, \Delta x^\mu \, \Delta x^\nu \; \quad \text{and} \quad \pi_\mu = p_\mu + \frac{i}{2} g_{\mu \nu}(x,y) \, \Delta x^\nu
\end{equation*}

\noindent one gets

\begin{multline}\label{TransitionAmplitudesRescaled}
    \langle x',y',\bar\eta \, | \, e^{-TH} \, | x,y,\eta \rangle = \Big( \sqrt{g(x,y)} \sqrt{g(x',y')} \Big)^{-1/2} \, e^{-\xi^2/4T} \, e^{\bar\eta \eta} \times \\
    \int \mathcal{D}p \mathcal{D}q \mathcal{D}\psi^\dagger \mathcal{D}\psi \; e^{-\int_0^1 dt \big\{ g^{\mu \nu}(x,y) \, p_{\mu} p_{\nu} + \Delta g^{\mu \nu} \pi_{\mu} \pi_{\nu} - i p_\mu \Dot{q}^\mu + \psi^\dagger \Dot{\psi} + T \, V_W \big\} } \; ,
\end{multline}

\noindent where $\Delta g^{\mu \nu}$ is evaluated according to \eqref{DeltaCurvature} and $V_W$ is evaluated at $\pi_\mu$, $x + t(x'-x) + \sqrt{T}q^1(t)$, $y + t(y'-y) + \sqrt{T}q^2(t)$, $\eta + \psi(t)$ and $\bar\eta + \psi^\dagger(t)$.

In order to turn \eqref{TransitionAmplitudesRescaled} into a computational efficient expression, it is useful to define the expectation value of any arbitrary functional of the fields as

\begin{multline}
    \Big\langle F[p,q,\psi,\psi^\dagger] \Big\rangle \\ = \frac{4 \pi T}{\sqrt{g(x,y)}} \int \mathcal{D}p \mathcal{D}q \mathcal{D}\psi^\dagger \mathcal{D}\psi \, e^{-\int_0^1 dt \big\{ g^{\mu \nu}(x,y) \, p_{\mu} p_{\nu} - i p_\mu \Dot{q}^\mu + \psi^\dagger \Dot{\psi} \big\}} F[p,q,\psi,\psi^\dagger] \; .
\end{multline}

\noindent The prefactor is chosen to fix the normalization according to $\langle 1 \rangle = 1$. Then \eqref{TransitionAmplitudesRescaled} turns into

\begin{equation}\label{TransitionAmplitudesExpectationValue}
    \langle x',y',\bar\eta \, | \, e^{-TH} \, | x,y,\eta \rangle = \frac{e^{\bar\eta \eta}}{4 \pi T} \bigg( \frac{g(x,y)}{g(x',y')} \bigg)^{1/4} \, e^{-\xi^2/4T} \bigg\langle e^{-\int_0^1 dt \big\{ \Delta g^{\mu \nu} p_{\mu} p_{\nu} + T \, V_W \big\} } \bigg\rangle \; .
\end{equation}

\noindent From the bosonic side, the expectation value in \eqref{TransitionAmplitudesExpectationValue} represents the phase space integration over trajectories which satisfy homogeneous Dirichlet conditions $q^\mu(0) = q^\mu(1) = 0$ in configuration space and no restriction at all in momentum space. This makes the relevant quadratic operator in the Gaussian measure of \eqref{TransitionAmplitudesExpectationValue} invertible. From the fermionic side, this expectation value represents the coherent-space integration over trajectories in terms of the ``coherent fields'' $\psi(t)$ and $\psi^\dagger(t)$ which satisfy $\psi(0) = \psi^\dagger(1) = 0$.

Finally, in order to compute the expectation values it is convenient to define the generating functional as

\begin{equation}\label{GeneratingFunctional1}
    Z[j,k,\Bar{\kappa},\kappa] \equiv \bigg\langle e^{\int_0^1 dt \big\{ i \, j_\mu q^\mu + i \, k^\mu p_\mu + \bar{\kappa} \psi - \psi^\dagger \kappa \big\}} \bigg\rangle \; ,
\end{equation}

\noindent where $j_\mu$, $k^\mu$, $\bar{\kappa}$ and $\kappa$ are ``six'' source fields ---of which $\bar{\kappa}$ and $\kappa$ are Grassmann fields. Completing squares one finds

\begin{multline}\label{GeneratingFunctional2}
    Z[j,k,\Bar{\kappa},\kappa] = \exp \bigg\{ \! - \! \int_0^1 \! dt \! \int_0^1 \! dt' \, \bigg( \frac{1}{4} g_{\mu \nu}(x,y) \, k^\mu(t) k^\nu(t') + i \, ^\bullet G(t,t') \, k^\mu(t) j_\mu(t') \\ + g^{\mu \nu}(x,y) \, G(t,t') \, j_\mu(t) j_\nu(t') + K(t,t') \, \bar\kappa(t) \kappa(t') \bigg) \bigg\} \; ,
\end{multline}

\noindent with the Green functions given by\footnote{While $G$ and $^\bullet \! G$ are different components of the Green matrix, the notation used ---following \cite{GreenBook}--- clarifies that $^\bullet \! G$ is equal to the derivative of $G$ with respect to the first argument. If it had coincided with the derivative with respect to the second argument, the notation $G^\bullet$ would have been used.}

\begin{align}
    G(t,t') & \equiv -\frac{1}{2}|t-t'| + \frac{1}{2}(t+t') - tt' \; , \\
    ^{\bullet}G(t,t') & \equiv -\frac{1}{2}\epsilon(t-t') + \frac{1}{2} - t' \; , \\
    K(t,t') & \equiv \frac{1}{2} \epsilon(t-t') + \frac{1}{2} \; ,
\end{align}

\noindent where $\epsilon(t-t')$ is the sign of $t-t'$ with the convention $\epsilon(0)=0$. Any expectation value of a functional that is polynomial in the integrated fields can be computed as functional derivatives of the generating functional in terms of the source fields, with the successive evaluation at zero source. Other sources give the expectation value for other specific functionals (for example, see \cite{CorradiniEdwardsHuetManzoPisani} for the computation of the expectation value of an exponential). In perturbative calculations, the relevant expectation values can usually be computed using Wick Theorem, for which only the two-point functions are needed. In the present case:

\begin{align}
        \big\langle p_\mu(t) \, p_\nu(t') \big\rangle & = \frac{1}{2} g_{\mu \nu}(x,y) \; ,\\
        \big\langle p_\mu(t) \, q^\nu(t') \big\rangle & = i \, ^\bullet G(t,t') \delta^\nu_\mu \; ,\\
        \big\langle q^\mu(t) \, q^\nu(t') \big\rangle & = 2 \, G(t,t') \, g^{\mu \nu}(x,y) \; ,\\
        \big\langle \psi(t) \, \psi^\dagger(t') \big\rangle & = K(t,t') \; ,
\end{align}

\noindent and every other two-point function is just zero.

\section{MIT bag boundary conditions}\label{section:MITbagBC}

In this section, a brief introduction to (Euclidean) MIT bag boundary conditions in a two-dimensional manifold $M$ is provided in terms of Dirac matrices as well as in terms of coherent states. Following Luckock \cite{Luckock1991}, consider the projector

\begin{equation}\label{MitBagProjector}
    \Pi_{-} = \frac{1}{2} \big( 1 - i \slashed{n} \gamma^{\text{ch}} \big)
\end{equation}

\noindent whose rank is equal to one-half of the total number of spinor components. For $D=2$, the rank is then equal to one. The MIT bag boundary conditions are given by

\begin{equation}\label{MitBagBoundaryCondition}
    \Pi_{-} \Psi |_{\partial M} = 0 \; .
\end{equation}

\noindent These local conditions are sufficient to ensure the self-adjointness of the Dirac operator $-i\slashed{D}$, which is a first-order differential operator. Equation \eqref{MitBagBoundaryCondition} imposes Dirichlet boundary conditions on one-half of the spinor components.

When working with the second-order differential operator $-\slashed{D}^2$, an additional consideration is made as usual: the relevant functional space is spanned by the eigenfunctions of the Dirac operator $-i\slashed{D}$. Therefore, not only does $\Psi$ obey the MIT bag boundary condition but so does $\slashed{D}\Psi$, implying $\Pi_{-}\slashed{D}\Psi|_{\partial M} = 0$. Commuting $\slashed{D}$ with $\Pi_{-}$ one finds

\begin{equation}\label{MitInducedBagBoundaryCondition}
    \bigg( n^\mu D_\mu - \frac{1}{2}L \bigg) \Pi_{+}\Psi |_{\partial M} = 0 \; ,
\end{equation}

\noindent where $L$ is the extrinsic curvature and

\begin{equation}\label{MitBagComplementaryProjector}
    \Pi_{+} = 1 - \Pi_{-} = \frac{1}{2} \big( 1 + i \slashed{n} \gamma^{\text{ch}} \big)
\end{equation}

\noindent is the projector complementary to $\Pi_{-}$. Hence, while the $\Pi_{-}$ projection of a spinor satisfies the ``Dirichlet-like'' boundary condition \eqref{MitBagBoundaryCondition}, the complementary $\Pi_{+}$ projection satisfies the ``Robin-like'' boundary condition \eqref{MitInducedBagBoundaryCondition}. When working with the Laplace-type operator $-\slashed{D}^2$, both \eqref{MitBagBoundaryCondition} and \eqref{MitInducedBagBoundaryCondition} are used, leading to a set of mixed boundary conditions. The total number of these boundary conditions now becomes exactly equal to the total number of spinor components.

The same approach can be repeated in terms of operators acting on coherent states rather than using Dirac matrices. Focusing on the manifold $M$ represented as a half-plane (whose boundary is the straight line $y=0$) and recalling the vielbein convention established in Section \ref{SectionGeometry}, one finds $\slashed{n} = g_{\mu \nu} \, e^\mu_i n^\nu \, \gamma^i = \gamma^2$. Thus, defining the difference between the complementary projectors as $\chi$, it follows

\begin{equation}\label{ChiDefinition}
    \chi \equiv \Pi_{+}-\Pi_{-} = i \slashed{n} \gamma^{\text{ch}} = -\gamma^1 \equiv -(\psi + \psi^\dagger) \; .
\end{equation}

\noindent Then the mixed MIT bag boundary conditions \eqref{MitBagBoundaryCondition} and \eqref{MitInducedBagBoundaryCondition} become

\begin{align}
    & \big( 1 + \psi + \psi^\dagger \big) | x,y,\eta \rangle \, \Big|_{y=0} = 0 \; ,\\
    & \bigg( n^\mu D_\mu - \frac{1}{2}L \bigg) \big( 1 - \psi - \psi^\dagger \big) | x,y,\eta \rangle \, \Big|_{y=0} = 0 \; .
\end{align}

\section{Heat-kernel in the manifold $M$}\label{MainSection}

This section summarises the main result of this article: a worldline representation for the heat-kernel of the Laplace-type operator $-\slashed{D}^2$ on the manifold $M$ expressed as a curved half-plane under MIT bag boundary conditions. Transition amplitudes \eqref{TransitionAmplitudesExpectationValue} in the manifold $\widetilde{M}$ without boundaries are used for this purpose.

Let

\begin{equation}\label{HeatKernelDefined}
    K_{M}(x,y,\eta;x',y',\bar{\eta};T) = \langle x',y',\bar\eta \, | \, e^{-T(-\slashed{D})^2} \, | x,y,\eta \rangle_{M}
\end{equation}

\noindent be the heat-kernel on $M$, (i.e. the transition amplitudes of a fermionic point-particle evolving with Hamiltonian $-\slashed{D}^2$). The ansatz proposed in the present article is

\begin{equation}\label{Ansatz}
    K_{M}(x,y,\eta;x',y',\bar{\eta};T) = \langle x',y',\bar\eta \, | \, e^{-T \widetilde{H}} \, | x,y,\eta \rangle + \langle x',-y',\bar\eta \, | \, \chi \, e^{-T \widetilde{H}} \, | x,y,\eta \rangle \; ,
\end{equation}

\noindent where the terms in the RHS represent transition amplitudes of a point-particle in $\widetilde{M}$ evolving with Hamiltonian

\begin{equation}\label{PointParticleHamiltonian}
    \widetilde{H}(x',y') = \big( \! -\slashed{D}^2 \big)^{\! >} (x',|y'|) + \Pi_{+} \sqrt{\frac{h}{g}} \, L \, \delta(y') \; .
\end{equation}

\noindent In this Hamiltonian, $\big( \! -\slashed{D}^2 \big)^{\! >}$ is given by \eqref{DiracSquaredAtUpperHalfPlane} and represents the Dirac operator squared $-\slashed{D}^2$ as in $\mathcal{H}(M)$, but symmetrized by being evaluated at $(x',|y'|)$. Note that since the RHS of \eqref{Ansatz} is defined in the entire double manifold $\widetilde{M}$, then the LHS is defined in $\widetilde{M}$ as well. Therefore, the heat-kernel in the LHS of \eqref{Ansatz} is, formally, an extension of the heat-kernel in the LHS of \eqref{HeatKernelDefined} from $M$ to $\widetilde{M}$. The first (second) term in the RHS of \eqref{Ansatz} is referred as a `direct' (`indirect') contribution to the heat-kernel.

To make sense of the ansatz \eqref{Ansatz}, first is necessary to specify what is meant when, for an arbitrary operator $\hat{O}$ defined as in \eqref{ReflectionOperators}, one writes $\hat{O}K_{M}$. It specifically means that the operator is acting on the final point of the heat-kernel:

\begin{equation}\label{OperatorsInAnsatz}
    \hat{O} K_{M}(x,y,\eta;x',y',\bar{\eta};T) = \langle x',y',\bar\eta \, |\hat{O}^> \, e^{-T \widetilde{H}} \, | x,y,\eta \rangle + \langle x',-y',\bar\eta \, |\hat{O}^< \, \chi \, e^{-T \widetilde{H}} \, | x,y,\eta \rangle \; .
\end{equation}

\noindent For instance, if one considers the projectors $\Pi_{\pm}$ (which satisfy $\Pi_{\pm}^> = \Pi_{\pm}^< = \Pi_{\pm}$) then

\begin{align}
    \Pi_{\pm} K_{M}(x,y,\eta;x',y',\bar{\eta};T) & = \pm \Pi_{\pm} K_{M}(x,y,\eta;x',-y',\bar{\eta};T) \; ,\label{HeatKernelProjectionSymmetry}\\
    \Pi_{\pm} \frac{\partial}{\partial y'} K_{M}(x,y,\eta;x',y',\bar{\eta};T) & = \mp \Pi_{\pm} \frac{\partial}{\partial y'} K_{M}(x,y,\eta;x',-y',\bar{\eta};T) \; .\label{HeatKernelDerivativeProjectionSymmetry}
\end{align}

\noindent This results are a direct consequence of the factor $\chi$ in the last term of \eqref{Ansatz}. Hence, the $\Pi_{-}$ ($\Pi_{+}$) projection of the heat-kernel is antisymmetric (symmetric) in terms of a reflection with respect to the boundary. This is a desired property of $K_M$, because previous research for scalar fields involving Dirichlet, Neumann and Robin boundary conditions \cite{BastianelliCorradiniPisani2007,BastianelliCorradiniPisaniSchubert,BastianelliCorradiniPisani2008,CorradiniEdwardsHuetManzoPisani,TrabajoDiploma} has shown that if Dirichlet (Robin) boundary conditions are satisfied, then the heat-kernel is antisymmetric (symmetric). Equation \eqref{PointParticleHamiltonian} is also inspired by these results for scalar fields: in the literature, the original Hamiltonian in $\mathcal{H}(M)$ has always been reflected symmetrically to the lower half-plane and, if Robin boundary conditions where obeyed, an additional delta term was introduced. The $\Pi_{+}$ projector accompanying the delta term in \eqref{PointParticleHamiltonian} ensures that this delta acts only on the projection that obeys a Robin boundary condition.

To verify if ansatz \eqref{Ansatz} is indeed equal to the heat-kernel \eqref{HeatKernelDefined} for $y' \geq 0$, it is necessary to check if \eqref{Ansatz} obeys the following conditions:

\begin{itemize}
    \item the heat equation in the bulk (that is, for $y' \neq 0$)

    \begin{equation}\label{HeatEquation}
        \bigg( -\slashed{D}^2 + \frac{d}{dT} \bigg) K_{M}(x,y,\eta;x',y',\bar{\eta};T) = 0 \; ,
    \end{equation}

    \item the Dirichlet boundary condition for the $\Pi_{-}$ projection

    \begin{equation}\label{DirichletCondition}
        \Pi_{-} K_{M}(x,y,\eta;x',y',\bar{\eta};T) \big|_{y'=0} = 0 \; ,
    \end{equation}

    \item the Robin boundary condition for the $\Pi_{+}$ projection

    \begin{equation}\label{RobinCondition}
        \bigg( n^\mu D_\mu - \frac{1}{2}L \bigg) \Pi_{+} K_{M}(x,y,\eta;x',y',\bar{\eta};T) \bigg|_{y'=0} = 0 \; ,
    \end{equation}

    \item and the initial condition in the bulk

    \begin{equation}\label{InitialCondition}
        K_{M}(x,y,\eta;x',y',\bar{\eta};T = 0) = \langle x' | x \rangle \langle y' | y \rangle \langle \bar\eta | \eta \rangle = e^{\bar\eta \eta} \, \delta(x-x') \delta(y-y') \; .
    \end{equation}
\end{itemize}

\noindent In the following, all the aforementioned requirements will be proved valid for \eqref{Ansatz}.

The Dirichlet condition \eqref{DirichletCondition} follows trivially from \eqref{HeatKernelProjectionSymmetry}. The initial condition \eqref{InitialCondition} is straightforward: from \eqref{Ansatz} it is trivial that $K_{M}(x,y,\eta;x',y',\bar{\eta};T = 0) = \langle x' | x \rangle \langle y' | y \rangle \langle \bar\eta | \eta \rangle + \langle x' | x \rangle \langle -y' | y \rangle \langle \bar\eta | \chi | \eta \rangle$, and since $y'>0$ in the bulk, one also finds $\langle -y' | y \rangle = \delta(y+y') = 0$ for every $y \geq 0$.

To prove the heat equation \eqref{HeatEquation}, expression \eqref{OperatorsInAnsatz} is used with $\hat{O} = -\slashed{D}^2$, which is given by \eqref{DiracSquaredReflected}. Then, considering the relation $(-\slashed{D}^2)^{<} \; \chi = \chi \; (-\slashed{D}^2)^{>}$ which follows from properties of the Dirac matrices, one gets

\begin{multline}\label{HeatEquationCheck1}
    -\slashed{D}^2 K_{M}(x,y,\eta;x',y',\bar{\eta};T) = \\
    \langle x',y',\bar\eta \, | \, (-\slashed{D}^2)^{>} \; e^{-T \widetilde{H}} \, | x,y,\eta \rangle + \langle x',-y',\bar\eta \, | \, \chi \; (-\slashed{D}^2)^{>} \; e^{-T \widetilde{H}} \, | x,y,\eta \rangle \; .
\end{multline}

\noindent On the other hand,

\begin{equation}\label{HeatEquationCheck2}
    -\frac{d}{dT} K_{M}(x,y,\eta;x',y',\bar{\eta};T) = \langle x',y',\bar\eta \, |\widetilde{H} \, e^{-T \widetilde{H}} \, | x,y,\eta \rangle + \langle x',-y',\bar\eta \, |\, \chi \, \widetilde{H} \, e^{-T \widetilde{H}} \, | x,y,\eta \rangle \; .
\end{equation}

\noindent From \eqref{HeatEquationCheck1} and \eqref{HeatEquationCheck2} it follows

\begin{equation}\label{HeatEquationEverywhere}
    \bigg( -\slashed{D}^2 + \frac{d}{dT} \bigg) K_{M}(x,y,\eta;x',y',\bar{\eta};T) = - \sqrt{\frac{h}{g}} \, L \, \delta(y') \, \Pi_{+} K_{M}(x,y,\eta;x',y',\bar{\eta};T) \; .
\end{equation}

\noindent Although \eqref{HeatEquationEverywhere} is valid in the entire manifold $\widetilde{M}$, note that its RHS is exactly zero in the bulk (where $y'>0$), from which \eqref{HeatEquation} is obtained.

The final requirement to prove is the Robin condition \eqref{RobinCondition}. Firstly, consider the integral of \eqref{HeatEquationEverywhere}, multiplied by $\sqrt{g}$, with respect to $y'$ from $0^-$ to $0^+$:

\begin{equation}\label{RobinCheck1}
    \int_{0^-}^{0^+} \! \! dy' \sqrt{g} \bigg( -\slashed{D}^2 + \frac{d}{dT} \bigg) K_{M}(x,y,\eta;x',y',\bar{\eta};T) + \sqrt{h} \, L \, \Pi_{+} K_{M}(x,y,\eta;x',y',\bar{\eta};T) \bigg|_{y'=0} = 0 \; .
\end{equation}

\noindent Next, note from equation \eqref{HeatKernelProjectionSymmetry} that $\Pi_{-} K_M$ is not necessarily continuous at the interface $y'=0$, but as long as the discontinuity is finite, the integral of $K_M$ is zero. The same conclusion is drawn for any derivative of $K_M$ with respect to any variable other than $y'$. Those observations hold even if $K_M$ or its derivatives are multiplied by any function with finite discontinuity. Therefore, \eqref{RobinCheck1} becomes

\begin{equation}\label{RobinCheck2}
    \int_{0^-}^{0^+} \! \! dy' D_2 \bigg( \sqrt{g} \frac{h}{g} D_2 \bigg) K_{M}(x,y,\eta;x',y',\bar{\eta};T) - \sqrt{h} \, L \, \Pi_{+} K_{M}(x,y,\eta;x',y',\bar{\eta};T) \bigg|_{y'=0} = 0 \; .
\end{equation}

\noindent This integral is solved easily if one considers the fact that $\omega_2|_{y'=0} = 0$ (see Section \ref{SectionGeometry}), the gauge semi-fixing condition \eqref{GaugeSemiFixing} and the symmetry property \eqref{HeatKernelDerivativeProjectionSymmetry}:

\begin{equation}\label{RobinCheck3}
    2 \frac{h}{\sqrt{g}} \frac{\partial}{\partial y'} \Pi_{+} K_{M}(x,y,\eta;x',y',\bar{\eta};T) \bigg|_{y'=0} - \sqrt{h} \, L \, \Pi_{+} K_{M}(x,y,\eta;x',y',\bar{\eta};T) \bigg|_{y'=0} = 0 \; .
\end{equation}

\noindent The Robin condition \eqref{RobinCondition} follows trivially from \eqref{RobinCheck3} and \eqref{NormalUnitVectorAtBoundary}.

\section{Application: Seeley-DeWitt coefficients}\label{section:SDW}

As an application of the previous construction, the first three Seeley-DeWitt coefficients $a_0$, $a_1$ and $a_2$ for the asymptotic heat-trace \eqref{HeatTraceAsymptoticExpansion} are computed in the present Section. In two dimensions the coefficient $a_2$ represents the (integrated) conformal anomaly of the theory.

The computation will be performed in phase space to avoid the introduction of ghost fields. Therefore, a phase space representation of the point-particle Hamiltonian \eqref{PointParticleHamiltonian} is needed. From the usual identification $\partial_\mu = ig^{1/4} \hat{p}_\mu \, g^{-1/4}$, introducing the counterterms \eqref{CountertermsGeneralEquation}, and taking Section \ref{section:CoherentStates} into consideration to work in terms of coherent states instead of Dirac matrices, one finds the Weyl ordered expression

\begin{equation}\label{HamiltonianPhaseSpace}
    \begin{aligned}
        \widetilde{H} = & \Big[ g^{\mu \nu} (\hat{p}_\mu + A_\mu)(\hat{p}_\nu + A_\nu) \Big]_S + \frac{1}{4} \omega^\mu \omega_\mu + \frac{1}{4} \Gamma^2 - \frac{1}{2} \Big( 1 + \psi + \psi^\dagger \Big) \, L \sqrt{\frac{h}{g}} \, \delta(y) \\
        & + 2 \, \Big\{ \Big[ \omega^\mu (\hat{p}_\mu + A_\mu) \Big]_S + g^{-1/2} \, F \Big\} \, (\psi \psi^\dagger)_A \; .
    \end{aligned}
\end{equation}

\noindent where the subscript $S$ stands for ``symmetrized'' with respect to the bosonic operators $\hat{p}_\mu$ and $\hat{x}^\mu$. The heat-trace then admits the following expression:

\begin{equation}
    \text{Tr }K_M = \text{Tr }K_M^{\text{dir}} + \text{Tr }K_M^{\text{ind}} \; ,
\end{equation}

\noindent with the `direct trace' $\text{Tr }K_M^{\text{dir}}$ and the `indirect trace' $\text{Tr }K_M^{\text{ind}}$ given by

\begin{align}
    \text{Tr }K_M^{\text{dir}} & = \int_M dx \, dy \sqrt{g} \int d\eta \, d\bar\eta \, e^{\bar\eta \eta} \langle x,y,\bar\eta \, | \, e^{-T \widetilde{H}} \, | x,y,\eta \rangle \; ,\label{DirectTraceDefinition}\\
    \text{Tr }K_M^{\text{ind}} & = \int_M dx \, dy \sqrt{g} \int d\eta \, d\bar\eta \, e^{\bar\eta \eta} \langle x,-y,\bar\eta \, | \, \chi \, e^{-T \widetilde{H}} \, | x,y,\eta \rangle \; .\label{IndirectTraceDefinition}
\end{align}

The computation for each contribution will be carried out in two steps: first, a general non-perturbative expression will be obtained by solving the Grassmann trace. Then, a perturbative calculation up to order $T^0$ will provide the SDW coefficients $a_0$, $a_1$ and $a_2$.

\subsection{Direct trace}

From Sections \ref{TransitionAmplitudesSection} and \ref{MainSection} it is now known that one can write \eqref{DirectTraceDefinition} as

\begin{equation}
    \text{Tr }K_M^{\text{dir}} = \frac{1}{4 \pi T} \int_M dx \, dy \sqrt{g} \int d\eta \, d\bar\eta \, e^{2\bar\eta \eta} \Big\langle e^{-\int_0^1 dt \, \widetilde{H}_\text{dir} } \Big\rangle
\end{equation}

\noindent where the exponent in the mean value is given by

\begin{equation}\label{DirectHamiltonianDefinition}
    \begin{aligned}
        \widetilde{H}_\text{dir} & = \Big[ g^{\mu \nu} \big( x+\sqrt{T}q^1, \big|y+\sqrt{T}q^2 \big| \big) - g^{\mu \nu}(x,y) \Big] p_\mu \, p_\nu + 2\sqrt{T} \, g^{\mu \nu} \, p_\mu \, A_\nu + T \, g^{\mu \nu} A_\mu A_\nu \\
        & \quad + \frac{T}{4} g^{\mu \nu} \omega_\mu \omega_\nu + \frac{T}{4} \Gamma^2 - \frac{T}{2} \Big( 1+ \psi + \psi^\dagger + \eta + \bar\eta \Big) L \sqrt{\frac{h}{g}} \, \delta \big( y + \sqrt{T}q^2 \big) \\
        & \quad + 2 \sqrt{T} \Big\{ g^{\mu \nu} \omega_\mu p_\nu + \sqrt{T} g^{\mu \nu} \omega_\mu A_\nu + \sqrt{T} g^{-1/2} F \Big\} \, (\psi \psi^\dagger + \psi \bar\eta + \eta \psi^\dagger + \eta \bar\eta ) \; .
    \end{aligned}
\end{equation}

\noindent Every function of the coordinates in \eqref{DirectHamiltonianDefinition} is evaluated at $\big( x+\sqrt{T}q^1, \big|y+\sqrt{T}q^2 \big| \big)$ unless explicitly stated. Note how $\widetilde{H}_\text{dir}$ is of order $\sqrt{T}$ at least, which eases the perturbative expansion in powers of $T$. Expanding $e^{-\int_0^1 dt \, \widetilde{H}_\text{dir} }$ in powers of $\eta$ and $\bar\eta$ and then solving the Grassmannian trace yields

\begin{multline}\label{DirectTraceNonPerturbative}
    \text{Tr }K_M^{\text{dir}} = \frac{1}{2 \pi T} \! \int_M \! \! dx \, dy \sqrt{g} \Big\langle e^{-\int_0^1 dt \, \mathcal{H}_\text{dir} } \\
    \times \Big( 1 + \! \sqrt{T} \, b_\text{dir} + 2T \, \theta_\text{dir} \bar\theta_\text{dir} - \frac{T^{3/2}}{2} (\theta_\text{dir} + \bar\theta_\text{dir}) \Delta_\text{dir} - \frac{T^2}{8}(\Delta_\text{dir})^2 \Big) \Big\rangle
\end{multline}

\noindent with

\begin{align}
    \mathcal{H}_\text{dir} & = \Big[ g^{\mu \nu} \big( x+\sqrt{T}q^1, \big|y+\sqrt{T}q^2 \big| \big) - g^{\mu \nu}(x,y) \Big] p_\mu \, p_\nu + 2\sqrt{T} \, g^{\mu \nu} \, p_\mu \, A_\nu + T \, g^{\mu \nu} A_\mu A_\nu \nonumber \\
    & \quad + \frac{T}{4} g^{\mu \nu} \omega_\mu \omega_\nu + \frac{T}{4} \Gamma^2 - \frac{T}{2} \Big( 1+ \psi + \psi^\dagger \Big) L \sqrt{\frac{h}{g}} \, \delta \big( y + \sqrt{T}q^2 \big) \nonumber \\
    & \quad + 2 \sqrt{T} \Big\{ g^{\mu \nu} \omega_\mu p_\nu + \sqrt{T} g^{\mu \nu} \omega_\mu A_\nu + \sqrt{T} g^{-1/2} F \Big\} \, \psi \psi^\dagger \; ,\label{DirectHamiltonianUseful}\\
    b_\text{dir} & = \int_0^1 dt \, \Big\{ g^{\mu \nu} \omega_\mu p_\nu + \sqrt{T} \, g^{\mu \nu} \omega_\mu A_\nu + \sqrt{T} \, g^{-1/2} F \Big\} \; ,\label{DirectBulk}\\
    \theta_\text{dir} & = \int_0^1 dt \, \Big\{ g^{\mu \nu} \omega_\mu p_\nu + \sqrt{T} \, g^{\mu \nu} \omega_\mu A_\nu + \sqrt{T} \, g^{-1/2} F \Big\} \psi \; ,\label{DirectTheta}\\
    \bar\theta_\text{dir} & = \int_0^1 dt \, \Big\{ g^{\mu \nu} \omega_\mu p_\nu + \sqrt{T} \, g^{\mu \nu} \omega_\mu A_\nu + \sqrt{T} \, g^{-1/2} F \Big\} \psi^\dagger \; ,\label{DirectBarTheta}\\
    \Delta_\text{dir} & = \int_0^1 dt \, L \sqrt{\frac{h}{g}} \, \delta \big( y + \sqrt{T}q^2 \big) \; .\label{DirectDelta}
\end{align}

\noindent In expressions \eqref{DirectHamiltonianUseful}-\eqref{DirectDelta} every function of the coordinates is again evaluated at $\big( x+\sqrt{T}q^1, \big|y+\sqrt{T}q^2 \big| \big)$ unless the evaluation point is explicitly stated. Equation \eqref{DirectTraceNonPerturbative} gives a non-perturbative result for the direct contribution of the heat-trace.

To compute the heat-trace up to any desired Seeley-DeWitt coefficient $a_k$, now one expands \eqref{DirectTraceNonPerturbative} in terms of $T$ up to order $T^{-1+k/2}$. Noticing that $\mathcal{H}_\text{dir}$ is of order $\sqrt{T}$ at least, this means that one needs to expand the exponential containing this factor up to order $k$ in the power series, thus ``lowering'' $\mathcal{H}_\text{dir}$ up to $k$ times in total. Then, defining

\begin{multline}
    \text{Tr }K_{M}^{\text{dir}(k)} = \frac{(-1)^k}{2 \pi T} \! \int_M \! \! dx \, dy \sqrt{g} \, \Big\langle \int_0^1 dt_1 \dots dt_k \, \mathcal{H}_\text{dir}(t_1) \dots \mathcal{H}_\text{dir}(t_k) \\ \times \Big( 1 + \! \sqrt{T} \, b_\text{dir} + 2T \, \theta_\text{dir} \bar\theta_\text{dir} - \frac{T^{3/2}}{2} (\theta_\text{dir} + \bar\theta_\text{dir}) \Delta_\text{dir} - \frac{T^2}{8}(\Delta_\text{dir})^2 \Big) \Big\rangle
\end{multline}

\noindent one can see that

\begin{equation*}
    \text{Tr }K_M^{\text{dir}} = \sum_{k=0}^\infty \frac{1}{k!} \, \text{Tr }K_{M}^{\text{dir}(k)} \; .
\end{equation*}

\noindent In particular, computing up to the coefficient $a_2$ means expanding up to order $T^0$, thus ``lowering'' $\mathcal{H}_\text{dir}$ up to $2$ times in total. Hence, one can readily write

\begin{equation}\label{DirectTraceExpansion}
    \text{Tr }K_M^{\text{dir}} = \text{Tr }K_M^{\text{dir}(0)} + \text{Tr }K_M^{\text{dir}(1)} + \frac{1}{2} \, \text{Tr }K_M^{\text{dir}(2)} + \mathcal{O}(T^{1/2}) \; .
\end{equation}

To compute $\text{Tr }K_M^{\text{dir}(0)}$, one first expands up to order $T^0$ obtaining

\begin{equation}\label{DirectZerothOrderPreliminaries}
    \text{Tr }K_{M}^{\text{dir}(0)} = \frac{1}{2 \pi T} \! \int_M \! \! dx \, dy \sqrt{g} \, \Big\langle 1 + \! \sqrt{T} \, b_\text{dir} + 2T \, \theta_\text{dir} \bar\theta_\text{dir} \Big\rangle + \mathcal{O}(T^{1/2}) \; ,
\end{equation}

\noindent where $b_\text{dir}$ is expanded up to order $T^{1/2}$, and in the product $\theta_\text{dir} \bar\theta_\text{dir}$ one only retains the lowest order which is $T^0$. In this scenario, using the mean values obtained at the end of Section \ref{TransitionAmplitudesSection}, one gets

\begin{align*}
    \langle b_\text{dir} \rangle & = \sqrt{T} \big( g^{\mu \nu} \omega_\mu A_\nu + g^{-1/2} F \big) + \mathcal{O}(T) \; ,\\
    \langle \theta_\text{dir} \bar\theta_\text{dir} \rangle & = \frac{1}{2} g^{\mu \nu} \omega_\mu \omega_\nu \int_0^1 dt dt' K(t,t') + \mathcal{O}(T^{1/2}) \; ,
\end{align*}

\noindent with the RHS of both equations evaluated at $(x,|y|)$. The integral of the Green function $K(t,t')$ yields $1/2$, so replacing into \eqref{DirectZerothOrderPreliminaries} one gets

\begin{equation}\label{DirectZerothOrder}
    \text{Tr }K_{M}^{\text{dir}(0)} = \frac{1}{2 \pi T} \text{Vol}(M) + \frac{1}{4 \pi} \int_M \! \! dx \, dy \sqrt{g} \bigg( g^{\mu \nu} \omega_\mu \omega_\nu + 2 g^{\mu \nu} \omega_\mu A_\nu + 2 g^{-1/2} F \bigg) + \mathcal{O}(T^{1/2}) \; .
\end{equation}

Moving on to $\text{Tr }K_M^{\text{dir}(1)}$ and expanding up to order $T^0$ one obtains

\begin{equation}\label{DirectFirstOrderPreliminaries1}
    \text{Tr }K_{M}^{\text{dir}(1)} = \frac{-1}{2 \pi T} \! \int_M \! \! dx \, dy \sqrt{g} \int_0^1 dt_1 \, \Big\langle \mathcal{H}_\text{dir}(t_1) \Big( 1 + \! \sqrt{T} \, b_\text{dir} \Big) \Big\rangle + \mathcal{O}(T^{1/2}) \; .
\end{equation}

\noindent Focusing first on the factor $1$ in the parenthesis, the mean value $\langle \mathcal{H}_\text{dir}(t_1) \rangle$ needs to be computed up to order $T$. To start this calculation, consider the ``$\delta$-term'' first:

\begin{multline}
    \int_M \! \! dx \, dy \sqrt{g} \int_0^1 \! \! dt_1 \bigg[ -\frac{T}{2} L \sqrt{\frac{h}{g}} \bigg\langle \Big( 1 + \psi + \psi^\dagger \Big) \, \delta\Big( y + \sqrt{T} q^2(t_1) \Big) \bigg\rangle \bigg] \\ = -\frac{T}{2} \int_M \! \! dx \, dy \sqrt{h} \, L \int_0^1 \! \! dt_1 \Big\langle \delta\Big( y + \sqrt{T} q^2(t_1) \Big) \Big\rangle \; .
\end{multline}

\noindent This expectation value can be computed by expressing the delta distribution in its Fourier representation:

\begin{equation*}
    \Big\langle \delta \Big( y + \sqrt{T} q^2(t_1) \Big) \Big\rangle = \int_{-\infty}^\infty \frac{dk}{2 \pi} e^{-iky}\Big\langle e^{ -ik\sqrt{T} \, q^2(t_1) } \Big\rangle \; .
\end{equation*}

\noindent Then the problem of computing the mean value of a delta turns into the problem of computing the mean value of an exponential, which can be achieved by using equations \eqref{GeneratingFunctional1} and \eqref{GeneratingFunctional2} with the source $j_\mu(t) = -\omega \, \delta_\mu^2 \, \delta(t-t_1)$ (see \cite{CorradiniEdwardsHuetManzoPisani} for a list with several mean values containing an exponential). This yields $\langle e^{ -ik\sqrt{T} \, q^2(t_1) } \rangle = e^{- T k^2 g^{22} G(t_1,t_1)}$. Then

\begin{equation*}
    \Big\langle \delta \Big( y + \sqrt{T} q^2(t_1) \Big) \Big\rangle = \bigg( \frac{g}{4 \pi T \, h \, G(t_1,t_1)} \bigg)^{\! \! 1/2} \, e^{-y^2 g/4T \, h \, G(t_1,t_1)} \; .
\end{equation*}

\noindent Inserting this into the integral on the manifold $M$ and rescaling $y \rightarrow \sqrt{4T} y$ yields

\begin{equation*}
    \int_M \! \! dx \, dy \sqrt{h} \, L \! \int_0^1 dt_1 \, \Big\langle \delta \Big( y + \sqrt{T}q^2(t_1) \Big) \Big\rangle = \frac{1}{\sqrt{\pi}} \! \int_M \! \! dx \, dy \sqrt{g} \, L \! \int_0^1 \! \frac{dt_1}{\sqrt{G(t_1,t_1)}} e^{-y^2 g/ h \, G(t_1,t_1)} \; ,
\end{equation*}

\noindent with the functions $g$ and $h$ on the RHS evaluated at $(x,\sqrt{4T}y)$. Expanding them in powers of $T$ leaves them evaluated at $(x,0)$ plus terms of order $T^{1/2}$. The integral in $y$ is then trivial at the lowest order:

\begin{equation*}
    \int_M \! \! dx \, dy \sqrt{h} \, L \! \int_0^1 dt_1 \, \Big\langle \delta \Big( y + \sqrt{T}q^2(t_1) \Big) \Big\rangle = \frac{1}{2} \int_{\partial M} \! \! dx \, \sqrt{h} \, L + \mathcal{O}(T^{1/2}) \; .
\end{equation*}

\noindent Now that the $\delta$-term in $\langle \mathcal{H}_\text{dir}(t_1) \rangle$ has been computed, one needs to solve the mean values of the other terms, each of them being proportional to some functions evaluated at $\big( x+\sqrt{T}q^1, \big|y+\sqrt{T}q^2 \big| \big)$. The general ideas about how to compute these mean values is left for Appendix \ref{AppendixSection}. The final result is

\begin{multline}\label{DirectFirstOrderPreliminaries2}
    \int_M \! \! dx \, dy \sqrt{g} \int_0^1 \! dt_1 \, \big\langle \mathcal{H}_\text{dir}(t_1) \big\rangle = \; \frac{T}{12} \int_M \! \! dx \, dy \sqrt{g} \bigg\{ g_{\mu \nu} g^{\rho \sigma} \partial_\rho \partial_\sigma g^{\mu \nu} - \partial_\mu \partial_\nu g^{\mu \nu} + 3 \Gamma^2 \\
    \begin{aligned}
    & + 12 \, g^{\mu \nu} A_\mu A_\nu + 3 \, g^{\mu \nu} \omega_\mu \omega_\nu + 12 \, g^{\mu \nu} \omega_\mu A_\nu + 12 g^{-1/2} F \bigg\} \\
    & - \frac{T}{12} \int_{\partial M} \! \! dx \, \sqrt{h} \, L + \mathcal{O}(T^{3/2}) \; .
    \end{aligned}
\end{multline}

Going back to \eqref{DirectFirstOrderPreliminaries1} and focusing now on the term with the factor $\sqrt{T} \, b_{\text{dir}}$ in the parenthesis, the mean value $\langle \mathcal{H}_\text{dir}(t_1) \, b_{\text{dir}} \rangle$ needs to be expanded up to order $T^{1/2}$. Since that is exactly the lowest order of $\mathcal{H}_\text{dir}$, then one only needs to retain the lowest order of both $\mathcal{H}_\text{dir}$ and $b_{\text{dir}}$. Through similar and less involved calculations as before, this results in

\begin{equation}\label{DirectFirstOrderPreliminaries3}
    \int_M \! \! dx \, dy \sqrt{g} \! \int_0^1 \! \! dt_1 \big\langle \mathcal{H}_\text{dir}(t_1) \, b_{\text{dir}} \big\rangle = \; \frac{\sqrt{T}}{2} \int_M \! \! dx \, dy  \sqrt{g} \bigg\{ 2 \, g^{\mu \nu} A_\mu \omega_\nu + \, g^{\mu \nu} \omega_\mu \omega_\nu \bigg\} + \mathcal{O}(T) \; .
\end{equation}

\noindent Inserting \eqref{DirectFirstOrderPreliminaries2} and \eqref{DirectFirstOrderPreliminaries3} into \eqref{DirectFirstOrderPreliminaries1} gives

\begin{multline}\label{DirectFirstOrder}
    \text{Tr }K_{M}^{\text{dir}(1)} = \; \frac{1}{24\pi} \int_M \! \! dx \, dy \sqrt{g} \bigg\{ -g_{\mu \nu} g^{\rho \sigma} \partial_\rho \partial_\sigma g^{\mu \nu} + \partial_\mu \partial_\nu g^{\mu \nu} - 3 \Gamma^2 \\
    \begin{aligned}
    & - 12 \, g^{\mu \nu} A_\mu A_\nu - 9 \, g^{\mu \nu} \omega_\mu \omega_\nu - 24 \, g^{\mu \nu} \omega_\mu A_\nu - 12 g^{-1/2} F \bigg\} \\
    & + \frac{1}{24 \pi} \int_{\partial M} \! \! dx \, \sqrt{h} \, L + \mathcal{O}(T^{1/2}) \; .
    \end{aligned}
\end{multline}

Only the term $\text{Tr }K_{M}^{\text{dir}(2)}$ remains to be computed. Its expansion up to order $T^0$ is

\begin{equation*}
    \text{Tr }K_{M}^{\text{dir}(2)} = \frac{1}{2 \pi T} \! \int_M \! \! dx \, dy \sqrt{g} \, \Big\langle \int_0^1 dt_1 dt_2 \, \mathcal{H}_\text{dir}(t_1) \mathcal{H}_\text{dir}(t_2) \Big\rangle + \mathcal{O}(T^{1/2}) \; .
\end{equation*}

\noindent Next, one needs to expand the product $\mathcal{H}_\text{dir}(t_1) \mathcal{H}_\text{dir}(t_2)$ up to order $T$, which is simply the lowest order. Through similar calculations as before:

\begin{multline}\label{DirectSecondOrder}
    \frac{1}{2} \text{Tr }K_{M}^{\text{dir}(2)} = \frac{1}{24 \pi} \! \int_M \! \! dx \, dy \sqrt{g} \bigg\{ \frac{1}{4} g^{\mu \nu} (\partial_\mu \log g) (\partial_\nu \log g) + (\partial_\mu g^{\mu \nu})(\partial_\nu \log g) \\
    \begin{aligned}
        & - \frac{1}{2} g^{\mu \nu} (\partial_\mu g_{\rho \sigma}) (\partial_\nu g^{\rho \sigma}) - g^{\mu \nu} (\partial_\rho g_{\mu \sigma}) (\partial_\nu g^{\rho \sigma}) \\
        & + 12 \, g^{\mu \nu} A_\mu A_\nu + 3 \, g^{\mu \nu} \omega_\mu \omega_\nu + 12 \, g^{\mu \nu} \omega_\mu A_\nu \bigg\} + \mathcal{O}(T^{1/2}) \; .
    \end{aligned}
\end{multline}

\noindent Adding equations \eqref{DirectZerothOrder}, \eqref{DirectFirstOrder} and \eqref{DirectSecondOrder}, and after some geometrical identities, yields

\begin{equation}\label{DirectTraceTotal}
    \text{Tr }K_{M}^{\text{dir}} = \frac{1}{2 \pi T} \text{Vol}(M) - \frac{1}{24 \pi} \int_M \! \! dx \, dy \sqrt{g} \, R + \frac{1}{24 \pi} \int_{\partial M} \! \! dx \, \sqrt{h} \, L + \mathcal{O}(T^{1/2}) \; .
\end{equation}

\subsection{Indirect trace}

The indirect trace is a bit more involved than the direct one for several reasons. First, note that the transition amplitude in \eqref{IndirectTraceDefinition} includes the term $\langle x,-y,\bar\eta \, | \, \chi$. Since the present model uses $\chi = -\psi - \psi^\dagger$, one can write

\begin{equation*}
    \langle x,-y,\bar\eta \, | \, \chi = -\bigg( \bar\eta + \frac{d}{d\bar\eta} \bigg) \langle x,-y,\bar\eta \, | \; .
\end{equation*}

\noindent This removes the operator $\chi$ in the transition amplitude at the expense of introducing the factor in parenthesis in the previous equation. Without $\chi$, the transition amplitude to be integrated is $\langle x,-y,\bar\eta \, | \, e^{-T \widetilde{H}} \, | x,y,\eta \rangle$, which can be expressed in terms of \eqref{TransitionAmplitudesExpectationValue}. The next complication is that now there is a multiplicative factor which is equal to $e^{-y^2g/hT}$. This term, when expanded as a power series, provides negative powers of $T$. To get rid of them, one can perform the rescaling $y \rightarrow \sqrt{T}y$.  With all these ingredients, equation \eqref{IndirectTraceDefinition} turns into

\begin{multline*}
    \text{Tr }K_M^{\text{ind}} = \frac{-1}{4 \pi \sqrt{T}} \int_M dx \, dy \sqrt{g(x,\sqrt{T}y)} \, e^{-y^2g(x,\sqrt{T}y)/h(x,\sqrt{T}y)} \\ \times \int d\eta \, d\bar\eta \, e^{\bar\eta \eta} \bigg( \bar\eta + \frac{d}{d\bar\eta} \bigg) e^{\bar\eta \eta} \Big\langle e^{-\int_0^1 dt \, \widetilde{H}_\text{ind} } \Big\rangle
\end{multline*}

\noindent where the exponent in the mean value is given by

\begin{equation}\label{IndirectHamiltonianDefinition}
    \begin{aligned}
        \widetilde{H}_\text{ind} & = \Big[ g^{\mu \nu} \big( x+\sqrt{T}q^1, \sqrt{T} \big|(1-2t)y+q^2 \big| \big) - g^{\mu \nu}(x,\sqrt{T}y) \Big] \pi_\mu \, \pi_\nu \\
        & \quad + 2\sqrt{T} \, g^{\mu \nu} \, \pi_\mu \, A_\nu + T \, g^{\mu \nu} A_\mu A_\nu \\
        & \quad + \frac{T}{4} g^{\mu \nu} \omega_\mu \omega_\nu + \frac{T}{4} \Gamma^2 - \frac{\sqrt{T}}{2} \Big( 1+ \psi + \psi^\dagger + \eta + \bar\eta \Big) L \sqrt{\frac{h}{g}} \, \delta \big( (1-2t)y + q^2 \big) \\
        & \quad + 2 \sqrt{T} \Big\{ g^{\mu \nu} \omega_\mu \pi_\nu + \sqrt{T} g^{\mu \nu} \omega_\mu A_\nu + \sqrt{T} g^{-1/2} F \Big\} \, (\psi \psi^\dagger + \psi \bar\eta + \eta \psi^\dagger + \eta \bar\eta ) \; ,
    \end{aligned}
\end{equation}

\noindent where $\pi_1 = p_1$ and $\pi_2 = p_2 - i \, y \, g(x,\sqrt{T}y) / h(x,\sqrt{T}y)$. Every function of the coordinates in \eqref{IndirectHamiltonianDefinition} is evaluated at $\big( x+\sqrt{T}q^1, \sqrt{T} \big|(1-2t)y+q^2 \big| \big)$ unless explicitly stated. $\widetilde{H}_\text{ind}$ is of order $\sqrt{T}$ at least, just like $\widetilde{H}_\text{dir}$. Expanding $e^{-\int_0^1 dt \, \widetilde{H}_\text{ind} }$ in powers of $\eta$ and $\bar\eta$ and then solving the Grassmannian trace yields

\begin{equation}\label{IndirectTraceNonPerturbative}
    \text{Tr }K_M^{\text{ind}} = \frac{1}{4 \pi} \! \int_M \! dx \, dy \sqrt{g(x,\sqrt{T}y)} \, e^{-y^2g(x,\sqrt{T}y)/h(x,\sqrt{T}y)} \Big\langle e^{-\int_0^1 dt \, \mathcal{H}_\text{ind} } \Big( 2(\bar\theta_\text{ind} - \theta_\text{ind}) - \Delta_\text{ind} \Big) \! \Big\rangle
\end{equation}

\noindent with

\begin{align}
    \mathcal{H}_\text{ind} & = \Big[ g^{\mu \nu} \big( x+\sqrt{T}q^1, \sqrt{T} \big|(1-2t)y+q^2 \big| \big) - g^{\mu \nu}(x,\sqrt{T}y) \Big] \pi_\mu \, \pi_\nu \nonumber \\
    & \quad + 2\sqrt{T} \, g^{\mu \nu} \, \pi_\mu \, A_\nu + T \, g^{\mu \nu} A_\mu A_\nu \nonumber \\
    & \quad + \frac{T}{4} g^{\mu \nu} \omega_\mu \omega_\nu + \frac{T}{4} \Gamma^2 - \frac{\sqrt{T}}{2} \Big( 1+ \psi + \psi^\dagger \Big) L \sqrt{\frac{h}{g}} \, \delta \big( (1-2t)y + q^2 \big) \nonumber \\
    & \quad + 2 \sqrt{T} \Big\{ g^{\mu \nu} \omega_\mu \pi_\nu + \sqrt{T} g^{\mu \nu} \omega_\mu A_\nu + \sqrt{T} g^{-1/2} F \Big\} \, \psi \psi^\dagger \; ,\label{IndirectHamiltonianUseful}\\
    \theta_\text{ind} & = \int_0^1 dt \, \Big\{ g^{\mu \nu} \omega_\mu \pi_\nu + \sqrt{T} \, g^{\mu \nu} \omega_\mu A_\nu + \sqrt{T} \, g^{-1/2} F \Big\} \psi \; ,\label{IndirectTheta}\\
    \bar\theta_\text{ind} & = \int_0^1 dt \, \Big\{ g^{\mu \nu} \omega_\mu \pi_\nu + \sqrt{T} \, g^{\mu \nu} \omega_\mu A_\nu + \sqrt{T} \, g^{-1/2} F \Big\} \psi^\dagger \; ,\label{IndirectBarTheta}\\
    \Delta_\text{ind} & = \int_0^1 dt \, L \sqrt{\frac{h}{g}} \, \delta \big( (1-2t)y + q^2 \big) \; .\label{IndirectDelta}
\end{align}

\noindent Once again, in expressions \eqref{IndirectHamiltonianUseful}-\eqref{IndirectDelta} every function of the coordinates is evaluated at $\big( x+\sqrt{T}q^1, \sqrt{T} \big|(1-2t)y + q^2 \big| \big)$ unless the evaluation point is explicitly stated. Equation \eqref{IndirectTraceNonPerturbative} gives a non-perturbative result for the indirect contribution of the heat-trace.

One interesting result that follows from equation \eqref{IndirectTraceNonPerturbative} is the fact that for boundaries $\partial M$ obeying $L=0$ one finds $\text{Tr }K_M^{\text{ind}} = 0$. Indeed, if $L=0$ then $\Delta_\text{ind} = 0$ and the mean value in the integrand becomes $\big\langle e^{-\int_0^1 dt \, \mathcal{H}_\text{ind} } (\bar\theta_\text{ind} - \theta_\text{ind}) \big\rangle$. Since $\bar\theta_\text{ind} - \theta_\text{ind}$ is linear in the fields $\psi$ and $\psi^\dagger$, it turns out that only those terms of $e^{-\int_0^1 dt \, \mathcal{H}_\text{ind}}$ that are also linear in the fields $\psi$ and $\psi^\dagger$ could give a non-zero mean value. But such terms do not exist if $L=0$, proving that $\text{Tr }K_M^{\text{ind}} = 0$ if the boundary has zero extrinsic curvature.

Another interesting fact lies in the aforementioned rescaling $y \rightarrow \sqrt{T} y$. Due to it, the evaluation point $(x,y)$ of any function is replaced by $(x,\sqrt{T} y)$. Hence, the coefficients in the subsequent expansion in powers of $T$ correspond to functions evaluated at $(x,0)$, that is, at the boundary. Therefore, indirect terms contribute only to boundary terms in the heat-trace expansion.

Note now that the RHS of equation \eqref{IndirectTraceNonPerturbative} is of order $T^0$ at least, which implies that the indirect terms do not contribute to the Seeley-DeWitt coefficients $a_0$ and $a_1$. For the coefficients $a_k$ with $k \geq 2$, one expands \eqref{IndirectTraceNonPerturbative} in terms of $T$ up to order $T^{-1+k/2}$. Note that $\mathcal{H}_\text{ind}$ is of order $\sqrt{T}$ at least, which means that one needs to expand the exponential containing this factor up to order $k-2$ in its power series, thus ``lowering'' $\mathcal{H}_\text{ind}$ up to $k-2$ times in total. In particular, computing up to the coefficient $a_2$ means that one can completely ignore the exponential containing the factor $\mathcal{H}_\text{ind}$. Therefore,

\begin{align*}
    \text{Tr }K_M^{\text{ind}} & = \frac{-1}{4 \pi} \! \int_M \! dx \, dy \sqrt{g(x,0)} \, e^{-y^2g(x,0)/h(x,0)} \Big\langle \Delta_\text{ind} \Big\rangle + \mathcal{O}(T^{1/2}) \\
    & = \frac{-1}{4 \pi} \! \int_M \! dx \, dy \sqrt{h(x,0)} \, L \, e^{-y^2g(x,0)/h(x,0)} \! \! \int_0^1 \! dt \Big\langle \delta \Big( (1-2t)y + q^2(t) \Big) \Big\rangle + \mathcal{O}(T^{1/2}) \; .
\end{align*}

\noindent The mean value of the delta is solved analogously as in the direct case. Thus one gets

\begin{equation}\label{IndirectTraceTotal}
    \text{Tr }K_M^{\text{ind}} = -\frac{1}{8 \pi} \int_{\partial M} \! dx \sqrt{h} \, L + \mathcal{O}(T^{1/2}) \; .
\end{equation}

\subsection{The trace anomaly}

Adding equations \eqref{DirectTraceTotal} and \eqref{IndirectTraceTotal} leads to the heat-trace on the manifold $M$:

\begin{equation*}
    \text{Tr }K_{M} = \frac{1}{2 \pi T} \text{Vol}(M) - \frac{1}{24 \pi} \int_M \! \! dx \, dy \sqrt{g} \, R - \frac{1}{12 \pi} \int_{\partial M} \! \! dx \, \sqrt{h} \, L + \mathcal{O}(T^{1/2}) \; .
\end{equation*}

\noindent By comparison with \eqref{HeatTraceAsymptoticExpansion}, the first three Seeley-DeWitt coefficients are obtained:

\begin{equation*}
    a_0 = 2 \, \text{Vol}(M) \; , \qquad \quad a_1 = 0 \; , \qquad \quad a_2 = - \frac{1}{6} \int_M \! \! dx \, dy \sqrt{g} \, R - \frac{1}{3} \int_{\partial M} \! \! dx \, \sqrt{h} \, L \; .
\end{equation*}

\noindent These results are in perfect agreement with \cite{HeatKernelManual, BransonGilkeyKirstenVassilevich}. In particular, the coefficient $a_2$ leads to the integrated trace anomaly:

\begin{equation}
    \int_M \! \! dx \, dy \, \sqrt{g} \, \langle T^\mu_{\; \; \mu} \rangle = \frac{a_2}{4 \pi} = - \frac{1}{24 \pi} \int_M \! \! dx \, dy \sqrt{g} \, R - \frac{1}{12 \pi} \int_{\partial M} \! \! dx \, \sqrt{h} \, L \; .
\end{equation}

\section{Conclusions and future work}\label{section:conclusion}

In this article, an approach to consistently implement the worldline formalism (WLF) for spinor particles in manifolds with boundaries was introduced. The boundary condition under consideration was the MIT bag boundary condition. To better fix ideas, the construction was carried out in an arbitrary two-dimensional manifold $M$ representable in half-plane coordinates $(x,y)$, such that the bulk of the manifold consists on $y>0$ and the boundary $\partial M$ is the straight line $y=0$. The construction relies on the method of images: the manifold was mirrored symmetrically with respect to the boundary, generating a new two-dimensional manifold $\widetilde{M}$ without boundaries. The old boundary became an interface were a singular curvature developed. To analyse the applicability of the method, the first three Seeley-DeWitt coefficients of the heat-trace were computed. In this approach, all curvature terms must be non-singular in the $(x,y)$ coordinate representation of the manifold.

The procedure that has been set up admits generalisations and concrete applications. Up to now, worldline formulations on quantum fields with boundaries have been established only for scalar fields: firstly, Dirichlet and Neumann boundary conditions were analysed in flat space using the method of images \cite{BastianelliCorradiniPisani2007}, and Robin boundary conditions were considered later using the same method \cite{BastianelliCorradiniPisaniSchubert}. Some years later, the method was extended to incorporate curved manifolds and curved boundaries, including compact manifolds \cite{CorradiniEdwardsHuetManzoPisani,TrabajoDiploma}. A different approach involving delta potentials, also for scalar fields, was analysed as well: firstly, a free scalar field with semitransparent Dirichlet boundary conditions was studied \cite{FranchinoPisani,FranchinoMazzitelli}, and semitransparent Neumann boundary conditions were analysed more recently \cite{AhmadiniazFranchinoManzoMazzitelli}. Shortly after, 
the case of semitransparent Dirichlet boundary conditions for scalar fields with a potential was studied using the same method \cite{Franchino}. None of the aforementioned worldline approaches considered, until now, spinor fields (but see \cite{Grosche1995,Grosche1999} for previous path integral approaches for Dirac particles in the presence of delta-like potentials). The procedure described in the present article is expected to open the path to include new scenarios, including new fields and new boundary conditions, in the context of the WLF.

For the case of new fields, the approach followed in the present article, as described above, could be regarded as splitting the projections of the field components according to whether they obey Dirichlet or Robin boundary conditions. The same procedure should be valid for any field obeying the same mixed boundary conditions regardless of the Lorentzian nature of the field. For instance, a multiplet of several scalar fields or a gauge (vector) field are scenarios where the present approach is expected to hold. For the latter case, physically meaningful mixed boundary conditions are present in the literature \cite{BransonGilkey,MossPoletti,Vassilevich1995,Vassilevich1998} (see also Section 3.4 in \cite{HeatKernelManual} for a brief summary). The implementation of the procedure described in the present article to gauge fields with these conditions is currently under consideration.

For the case of other boundary conditions, it is well known that spinors could satisfy a large variety of them. In the area of local boundary conditions, MIT bag conditions are just one of many possibilities. Another closely related, physically meaningful possibility are the chiral bag boundary conditions, first introduced in \cite{ChodosThorn} to construct a chirally symmetric theory in Quantum Chromodynamics. Since then, they were applied in other areas such as fermionic monopoles \cite{Yamagishi}, fermionic billiards \cite{BerryMondragon}, graphene devices \cite{ZhangDietz} and Weyl semimetals \cite{IvanovKurkovVassilevich}. These conditions can be written as $\Pi_{-} \Psi|_{\partial M} = 0$ using the non-Hermitian projector $\Pi_{-} = \frac{1}{2} \big( 1 - i \slashed{n} \gamma^{\text{ch}} \, e^{r(x^\alpha) \gamma^{\text{ch}}} \big)$, where $r(x^\alpha)$ is a real-valued function of the tangential coordinates. The main difficulty with these conditions is that, for $r(x^\alpha) \neq 0$, the first-order induced boundary condition contains partial derivatives \cite{EspositoKirsten}. Furthermore, first-order boundary conditions with tangential derivatives are completely unexplored within the worldline frame. It would be interesting to see if one could implement a worldline procedure, similar to the one of the present article, to include these boundary conditions. Besides, they are reduced to MIT bag boundary conditions for $r(x^\alpha) = 0$, so the present article could be use to check the construction of any possible procedure.

In the area of global boundary conditions, the Atiyah-Patodi-Singer (APS) conditions, sometimes referred as spectral boundary conditions, are well known in the literature \cite{AtiyahPatodiSinger,Abrikosov}. They are connected to index theorems \cite{Witten,DabholkarJainRudra,FukayaFurutaMatsuoOnogiYamaguchi,KobayashiYonekura} and raised interest in the study of topological devices \cite{AsoreyBalachandranPerezPardo}. The Seeley-DeWitt coefficients of the heat-kernel asymptotics for these boundary conditions have been studied in the past \cite{DowkerGilkeyKirsten,GilkeyKirsten}. They can be written in terms of projectors $\Pi_{\pm}$, but they are usually expressed in terms of eigenvector components which are not perfectly suited for worldline computations. It would be interesting to see if APS boundary conditions can be implemented within a worldline perspective, where the familiar calculational efficiency of worldline techniques is expected to hold.

Besides the aforementioned generalisations, concrete applications are also at hand. On the one hand, fermions in manifolds with boundaries have been used to study quantities in Quantum Field Theories such as anomalies \cite{FoscoSilva} and the Casimir effect \cite{SundbergJaffe}. These approaches are based on delta-like potentials instead of the method of images. On the other hand, worldline numerical computations have been employed on the same areas \cite{GiesLangfeldMoyaerts,EdwardsGonzalezDominguezHuetTrejo,GiesKlingmuller,EdwardsGerberSchubertTrejoWeber,EdwardsGerberSchubertTrejoTsiftsiWeber}, mostly considering rigid boundaries and scalar fields. The approach provided in the present article could be used to study these phenomena analytically for spinor fields, as well to test numerical computations in manifolds with rigid boundaries.

\acknowledgments

The author thanks Santiago Christiansen Murguizur, Horacio Falomir, Joaquín Liniado and Pablo Pisani for valuable discussions during the research. This work was supported by CONICET (under the program ``Becas Internas Doctorales'', RESOL-2020-129-APN-DIR\#CONICET) and UNLP (under project 11/X909).

\appendix

\section{Mean value of symmetrically reflected functions}\label{AppendixSection}

Consider an arbitrary function $f(x,y)$ that is finite at $y=0$. If one replaces $y \rightarrow |y|$ it is possible to write $f(x,|y|) = f(x,y) - 2 \theta(-y) f_A(x,y)$ with $f_A(x,y) \equiv \big\{ f(x,y) - f(x,-y) \big\}/2$. Consider then the problem of computing the following mean value:

\begin{multline}\label{Appendix1}
    \Big\langle f \big(x+ \sqrt{T} q^1, |y+\sqrt{T}q^2| \big) \mathcal{M} \Big\rangle = \Big\langle f \big(x+ \sqrt{T} q^1 , y+\sqrt{T}q^2 \big) \mathcal{M} \Big\rangle \\ - 2 \, \Big\langle \theta \big( -y -\sqrt{T}q^2 \big)  \, f_A \big(x+ \sqrt{T} q^1 , y+\sqrt{T}q^2 \big) \mathcal{M} \Big\rangle \; ,
\end{multline}

\noindent which appears frequently in the calculation of the direct trace. $\mathcal{M}$ represents any monomial of the fields $q^\mu$ and $p_\mu$. In the following it will be shown how to compute \eqref{Appendix1} based on the ideas set in \cite{CorradiniEdwardsHuetManzoPisani}.

To solve the mean value with the step function, perform the Fourier transformation

\begin{equation*}
    \theta(-y) = \! \! \int_{-\infty}^\infty \frac{dk}{2 \pi i} \frac{e^{-i k y}}{k - i0} = \! \! \int_{-\infty}^\infty dk \, \frac{e^{-i k y}}{2 \pi i} \bigg[ \text{PV} \bigg( \frac{1}{k} \bigg) + i\pi \delta( k ) \bigg] \; .
\end{equation*}

\noindent Then \eqref{Appendix1} becomes

\begin{multline}\label{Appendix2}
    \Big\langle f \big(x+ \sqrt{T} q^1, |y+\sqrt{T}q^2| \big) \mathcal{M} \Big\rangle = \Big\langle f \big(x+ \sqrt{T} q^1 , y+\sqrt{T}q^2 \big) \mathcal{M} \Big\rangle \\ - \int_{-\infty}^\infty \! \! dk \, \frac{e^{-i k y/\sqrt{T}}}{i \pi} \bigg[ \text{PV} \bigg( \frac{1}{k} \bigg) + i\pi \delta( k ) \bigg] \Big\langle e^{-i k q^2}  \, f_A \big(x+ \sqrt{T} q^1 , y+\sqrt{T}q^2 \big) \mathcal{M} \Big\rangle \; .
\end{multline}

\noindent The computation of the mean values in the RHS of \eqref{Appendix2} can now be carried out by expanding $f$ and $f_A$ as power series with respect to the fields $q^\mu$. The mean value with the exponential is computed employing the generating functional given by \eqref{GeneratingFunctional1} and \eqref{GeneratingFunctional2} with $j_\mu = -k \, \delta^2_\mu \, \delta(t-t')$, where $t'$ is the time at which the field $q^2(t')$ in the exponent is evaluated. An explicit list of several mean values involving this exponential is given in \cite{CorradiniEdwardsHuetManzoPisani}.

In calculations of the heat-trace, equation \eqref{Appendix2} appears integrated in the manifold $M$ with measure $\sqrt{g}$. In this scenario, introducing the rescaling $y \rightarrow \sqrt{T} y$ yields

\begin{multline}\label{Appendix3}
    \int_M \! \! dx \, dy \sqrt{g} \Big\langle f \big(x+ \sqrt{T} q^1, |y+\sqrt{T}q^2| \big) \mathcal{M} \Big\rangle = \int_M \! \! dx \, dy \sqrt{g} \, \Big\langle f \big(x+ \sqrt{T} q^1 , y+\sqrt{T}q^2 \big) \mathcal{M} \Big\rangle \\ - \sqrt{T} \! \! \int_M \! \! dx \, dy \sqrt{g(x,\sqrt{T}y)} \! \! \int_{-\infty}^\infty \! \! \!  dk \, \frac{e^{-i k y}}{i \pi} \bigg[ \text{PV} \bigg( \frac{1}{k} \bigg) \! \! + \! i\pi \delta( k ) \bigg] \\ \times \Big\langle e^{-i k q^2} f_A \big(x \! + \! \sqrt{T} q^1 , \sqrt{T} (y \! + \! q^2) \big) \mathcal{M} \Big\rangle_{y \rightarrow \sqrt{T} y} ,
\end{multline}

\noindent where the result of the mean value in the third line should be evaluated at $(x,\sqrt{T} y)$ due to the rescaling. Note that the last term (second and third line) in the RHS is at least of order $T$: indeed, the lowest explicit order is $\sqrt{T}$, but its coefficient is proportional to $f_A(x,0) = 0$.

Several particular applications of equation \eqref{Appendix3} are of interest for the calculations of Section \ref{section:SDW}. As an example, consider the monomial $\mathcal{M} = p_\mu p_\nu$ and $f = g^{\mu \nu}$. Up to order $T$, equation \eqref{Appendix3} yields

\begin{multline}\label{Appendix4}
    \int_M \! \! dx \, dy \sqrt{g} \, \Big\langle  \Big[ g^{\mu \nu} \Big(x \! + \! \sqrt{T} q^1, |y \! + \! \sqrt{T}q^2| \Big) - g^{\mu \nu}(x,y) \Big] p_\mu p_\nu \Big\rangle \\
    \begin{aligned}
        & = \frac{T}{2} \! \int_M \! \! \! dx \, dy \sqrt{g} \Big\{ g_{\mu \nu} g^{\rho \sigma} \partial_\rho \partial_\sigma g^{\mu \nu} G(t,t) - 2 \partial_\mu \partial_\nu g^{\mu \nu} \, ^\bullet G(t,t)^2 \Big \} \\
        & \quad + \frac{T}{2} \int_{-\infty}^\infty dx \sqrt{g} \, \partial_2 g^{\mu \nu} \Big\{ g^{22}g_{\mu \nu} \, G(t,t) - 2 \delta_\mu^2 \delta_\nu^2 \, ^\bullet G(t,t)^2 \Big\} \Big|_{y=0} + \mathcal{O}(T^{3/2}) \; ,
    \end{aligned}
\end{multline}

\noindent where each curvature element in the integral of the last line is being evaluated at $(x,0)$.

As a final remark, the present procedure to compute mean values can be extended to the case of products of several functions $\langle f_1 f_2 \dots f_l \rangle$, where each function $f_i$ is being evaluated at $\big( x + \sqrt{T} q^1(t_i) , \big| y + \sqrt{T} q^2(t_i) \big| \big)$. For each of them it is possible to write, as before, a ``bulk term'' plus a ``$\theta$-term'', with its own Fourier transformation for each step function.



\begin{thebibliography}{99}

\bibitem{BernKosower}
Z. Bern and D. A. Kosower, \emph{Efficient calculation of one loop QCD amplitudes}, \emph{Phys. Rev. Lett.} {\bf 66} (1991) 1669.

\bibitem{Strassler}
M. J. Strassler, \emph{Field theory without Feynman diagrams: One loop effective
actions}, \emph{Nucl. Phys. B} {\bf 385} (1992) 145 [\href{https://arxiv.org/abs/hep-ph/9205205}{\texttt{arXiv:hep-ph/9205205}}].

\bibitem{Schubert2001}
C. Schubert, \emph{Perturbative quantum field theory in the string-inspired formalism}, \emph{Phys. Rept.} {\bf 355} (2001) 73 [\href{https://arxiv.org/abs/hep-th/0101036}{\texttt{arXiv:hep-th/0101036}}].

\bibitem{BastianelliZirotti}
F. Bastianelli and A. Zirotti, \emph{Worldline formalism in a gravitational background}, \emph{Nucl. Phys. B} {\bf 642} (2002) 372 [\href{https://arxiv.org/abs/hep-th/0205182}{\texttt{arXiv:hep-th/0205182}}].

\bibitem{BastianelliCorradiniZirotti}
F. Bastianelli, O. Corradini and A. Zirotti, \emph{Dimensional regularization for $\mathcal{N} = 1$ supersymmetric sigma models and the worldline formalism}, \emph{Phys. Rev. D} {\bf 67} (2003) 104009 [\href{https://arxiv.org/abs/hep-th/0211134}{\texttt{arXiv:hep-th/0211134}}].

\bibitem{BastianelliBenincasaGiombi}
F. Bastianelli, P. Benincasa and S. Giombi, \emph{Worldline approach to vector and antisymmetric tensor fields}, \emph{JHEP} {\bf 04} (2005) 010 [\href{https://arxiv.org/abs/hep-th/0503155}{\texttt{arXiv:hep-th/0503155}}].

\bibitem{BastianelliCorradiniLatini}
F. Bastianelli, O. Corradini and E. Latini, \emph{Spinning particles and higher spin fields on (A)dS backgrounds}, \emph{JHEP} {\bf 11} (2008) 054 [\href{https://arxiv.org/abs/0810.0188}{\texttt{arXiv:0810.0188}}].

\bibitem{Corradini2010}
O. Corradini, \emph{Half-integer higher spin fields in (A)dS from spinning particle models}, \emph{JHEP} {\bf 09} (2010) 113 [\href{https://arxiv.org/abs/1006.4452}{\texttt{arXiv:1006.4452}}].

\bibitem{BastianelliBonezziCorradiniLatini}
F. Bastianelli, R. Bonezzi, O. Corradini and E. Latini, \emph{Effective action for higher spin fields on (A)dS backgrounds}, \emph{JHEP} {\bf 12} (2012) 113 [\href{https://arxiv.org/abs/1210.4649}{\texttt{arXiv:1210.4649}}].

\bibitem{GiesLangfeldMoyaerts}
H. Gies, K. Langfeld and L. Moyaerts, \emph{Casimir effect on the worldline}, \emph{JHEP} {\bf 06} (2003) 018 [\href{https://arxiv.org/abs/hep-th/0303264}{\texttt{arXiv:hep-th/0303264}}].

\bibitem{FoscoLombardoMazzitelli}
C. D. Fosco, F. C. Lombardo and F. D. Mazzitelli, \emph{Neumann Casimir effect:
a singular boundary-interaction approach}, \emph{Phys. Lett. B} {\bf 690} (2010) 189 [\href{https://arxiv.org/abs/0912.0886}{\texttt{arXiv:0912.0886}}].

\bibitem{EdwardsGonzalezDominguezHuetTrejo}
J. P. Edwards, V. A. González-Domínguez, I. Huet, M. A. Trejo, \emph{Non-perturbative Quantum Propagators in Bounded Spaces}, \emph{In review} [\href{https://arxiv.org/abs/2110.04969}{\texttt{arXiv:2110.04969}}].

\bibitem{FranchinoMazzitelli}
S. A. Franchino-Viñas and F. D. Mazzitelli, \emph{Effective action for delta potentials: Spacetime-dependent inhomogeneities and Casimir self-energy}, \emph{Phys. Rev. D} {\bf 103} (2021) 065006 [\href{https://arxiv.org/abs/2010.11144}{\texttt{arXiv:2010.11144}}].

\bibitem{AhmadiniazFranchinoManzoMazzitelli}
N. Ahmadiniaz, S. A. Franchino-Viñas, L. Manzo and F. D. Mazzitelli, \emph{Local Neumann semitransparent layers: resummation, pair production and duality}, \emph{Phys. Rev. D} {\bf 106} (2022) 105022 [\href{https://arxiv.org/abs/2208.07383}{\texttt{arXiv:2208.07383}}].

\bibitem{Franchino}
S. A. Franchino-Viñas, \emph{Resummed heat-kernel and form factors for surface contributions: Dirichlet semitransparent boundary conditions}, \emph{J. Phys. A: Math. Theor.} {\bf 56} (2023) 115202 [\href{https://arxiv.org/abs/2208.11979}{\texttt{arXiv:2208.11979}}].

\bibitem{FranchinoPisani}
S. A. Franchino-Viñas and P. Pisani, \emph{Semi-transparent Boundary Conditions in the Worldline Formalism}, \emph{J. Phys. A} {\bf 44} (2011) 295401 [\href{https://arxiv.org/abs/1012.2883}{\texttt{arXiv:1012.2883}}].

\bibitem{BastianelliCorradiniPisani2007}
F. Bastianelli, O. Corradini and P. Pisani, \emph{Worldline approach to quantum field theories on flat manifolds with boundaries}, \emph{JHEP} {\bf 02} (2007) 059 [\href{https://arxiv.org/abs/hep-th/0612236}{\texttt{arXiv:hep-th/0612236}}].

\bibitem{BastianelliCorradiniPisaniSchubert}
F. Bastianelli, O. Corradini, P. Pisani and C. Schubert, \emph{Scalar heat kernel with boundary in the worldline formalism}, \emph{JHEP} {\bf 10} (2008) 095 [\href{https://arxiv.org/abs/0809.0652}{\texttt{arXiv:0809.0652}}].

\bibitem{BastianelliCorradiniPisani2008}
F. Bastianelli, O. Corradini and P. Pisani, \emph{Scalar field with Robin boundary conditions in the worldline formalism}, \emph{J. Phys. A} {\bf 41} (2008) 164010 [\href{https://arxiv.org/abs/0710.4026}{\texttt{arXiv:0710.4026}}].

\bibitem{CorradiniEdwardsHuetManzoPisani}
O. Corradini, J. P. Edwards, I. Huet, L. Manzo and P. Pisani, \emph{Worldline formalism for a confined scalar field}, \emph{JHEP} {\bf 08} (2019) 037 [\href{https://arxiv.org/abs/1905.00945}{\texttt{arXiv:1905.00945}}].

\bibitem{TrabajoDiploma}
L. Manzo, \emph{Formalismo línea de mundo para un campo escalar en variedades curvas y con borde}, (\emph{Undergraduate Thesis}) (2020) [\href{http://sedici.unlp.edu.ar/handle/10915/116833}{\texttt{sedici.unlp.edu.ar/handle/10915/116833}}].

\bibitem{ChodosJaffeJohnsonThornWeisskopf}
A. Chodos, R. L. Jaffe, K. Johnson, C. B. Thorn and V. F. Weisskopf, \emph{New extended model of hadrons}, \emph{Phys. Rev. D} {\bf 9} (1974) 3471.

\bibitem{BreitenlohnerFreedman}
P. Breitenlohner and D. Z. Freedman, \emph{Stability in Gauged Extended Supergravity}, \emph{Ann. Phys.} {\bf 144} (1982) 249.

\bibitem{BerryMondragon}
M. V. Berry and R. J. Mondragón, \emph{Neutrino billiards: time reversal
symmetry-breaking without magnetic fields}, \emph{Proc. R. Soc. Lond. A} {\bf 412} (1987) 53.

\bibitem{SheikhJabbari}
M. M. Sheikh-Jabbari, \emph{More on mixed boundary conditions and D-branes bound states}, \emph{Phys. Lett. B} {\bf 425} (1998) 48 [\href{https://arxiv.org/abs/hep-th/9712199}{\texttt{arXiv:hep-th/9712199}}].

\bibitem{LuckockMoss}
H. Luckock and I. Moss, \emph{The quantum geometry of random surfaces and
spinning membranes}, \emph{Class. Quantum Grav.} {\bf 6} (1989) 1993.

\bibitem{BeneventanoSantangelo}
C. G. Beneventano and E. M. Santangelo, \emph{Boundary conditions in the
Dirac approach to graphene devices}, \emph{Int. J. Mod. Phys.: Conf. Ser.} {\bf 14} (2012) 240 [\href{https://arxiv.org/abs/1011.2772}{\texttt{arXiv:1011.2772}}].

\bibitem{GreenBook}
F. Bastianelli and P. van Nieuwenhuizen, \emph{Path integrals and anomalies in
curved space}, Cambridge University Press (2009).

\bibitem{Schwinger}
J. Schwinger, \emph{On Gauge Invariance and Vacuum Polarization}, \emph{Phys. Rev.} {\bf 82} (1951) 664.

\bibitem{Schrodinger}
E. Schrödinger, \emph{Diracsches Elektron im Schwerefeld I}, \emph{Sitz. Preuss. Akad. Wiss. Berlin} (1932) 105.

\bibitem{Lichnerowicz}
A. Lichnerowicz, \emph{Spineurs harmoniques}, \emph{C. R. Acad. Sci. Paris} {\bf 257} (1963) 7.

\bibitem{HeatKernelManual}
D. V. Vassilevich, \emph{Heat kernel expansion: User’s manual}, \emph{Phys. Rept.} {\bf 388} (2003) 279 [\href{https://arxiv.org/abs/hep-th/0306138}{\texttt{arXiv:hep-th/0306138}}].

\bibitem{Fradkin}
E. S. Fradkin, \emph{Application of functional methods in quantum field theory and quantum statistics (II)}, \emph{Nucl. Phys.} {\bf 76} (1966) 588.

\bibitem{BerezinMarinov}
F. A. Berezin and M. S. Marinov, \emph{Particle Spin Dynamics as the Grassmann Variant of Classical Mechanics}, \emph{Ann. Phys.} {\bf 104} (1977) 336.

\bibitem{FradkinGitman}
E. S. Fradkin and D. M. Gitman, \emph{Path-integral representation for the relativistic particle propagators and BFV quantization}, \emph{Phys. Rev. D} {\bf 44} (1991) 3230.

\bibitem{HenneauxTeitelboim}
M. Henneaux and C. Teitelboim, \emph{quantization of Gauge Systems}, Princeton University Press (1992).

\bibitem{AhmadiniazBandaGuzmanBastianelliCorradiniEdwardsSchubert}
N. Ahmadiniaz, V. M. Banda Guzmán, F. Bastianelli, O. Corradini, J. P. Edwards and C. Schubert, \emph{Worldline master formulas for the dressed electron propagator. Part I. Off-shell amplitudes}, \emph{JHEP} {\bf 08} (2020) 049 [\href{https://arxiv.org/abs/2004.01391}{\texttt{arXiv:2004.01391}}].

\bibitem{OhnukiKashiwa}
Y. Ohnuki, T. Kashiwa, \emph{Coherent States of Fermi Operators and the Path Integral}, \emph{Prog. Theor. Phys.} {\bf 60} (1978) 548.

\bibitem{BordiCasalbuoni}
F. Bordi and R. Casalbuoni, \emph{Dirac Propagator from Path Integral quantization of the Pseudoclassical Spinning Particle}, \emph{Phys. Lett. B} {\bf 93} (1980) 308.

\bibitem{HentyHoweTownsend}
J. C. Henty, P. S. Howe and P. K. Townsend, \emph{Quantum Mechanics of the Relativistic Spinning Particle}, \emph{Class. Quant. Grav.} {\bf 5} (1988) 807.

\bibitem{vanHolten}
J. W. van Holten, \emph{Propagators and Path Integrals}, \emph{Nucl. Phys. B} {\bf 457} (1995) 375 [\href{https://arxiv.org/abs/hep-th/9508136}{\texttt{arXiv:hep-th/9508136}}].

\bibitem{DHokerGagne1}
E. D'Hoker and D. G. Gagné, \emph{Worldline Path Integrals for Fermions with Scalar, Pseudoscalar and Vector Couplings}, \emph{Nucl. Phys. B} {\bf 467} (1996) 272 [\href{https://arxiv.org/abs/hep-th/9508131}{\texttt{arXiv:hep-th/9508131}}].

\bibitem{DHokerGagne2}
E. D'Hoker and D. G. Gagné, \emph{Worldline Path Integrals for Fermions with General Couplings}, \emph{Nucl. Phys. B} {\bf 467} (1996) 297 [\href{https://arxiv.org/abs/hep-th/9512080}{\texttt{arXiv:hep-th/9512080}}].

\bibitem{Bhattacharya}
S. Bhattacharya, \emph{Worldline Path-Integral Representations for Standard Model Propagators and Effective Actions}, \emph{Adv. High Energy Phys.} {\bf 2017} (2017) 2165731.

\bibitem{Luckock1991}
H. Luckock, \emph{Mixed boundary conditions in quantum field theory}, \emph{J.
Math. Phys.} {\bf 32} (1991) 1775.

\bibitem{BransonGilkeyKirstenVassilevich}
T. P. Branson, P. B. Gilkey, K. Kirsten and D. V. Vassilevich, \emph{Heat kernel asymptotics with mixed boundary conditions}, \emph{Nucl. Phys. B} {\bf 563} (1999) 603 [\href{https://arxiv.org/abs/hep-th/9906144}{\texttt{arXiv:hep-th/9906144}}].

\bibitem{Grosche1995}
C. Grosche, \emph{Delta-prime function perturbations and Neumann boundary conditions by path integration}, \emph{J. Phys. A: Math. Gen.} {\bf 28} (1995) L99 [\href{https://arxiv.org/abs/hep-th/9402110}{\texttt{arXiv:hep-th/9402110}}].

\bibitem{Grosche1999}
C. Grosche, \emph{Boundary conditions in path integrals from point interactions for the path integral of the one-dimensional Dirac particle}, \emph{J. Phys. A: Math. Gen.} {\bf 32} (1999) 1675.

\bibitem{BransonGilkey}
T. P. Branson and P. B. Gilkey, \emph{The asymptotics of the Laplacian on a manifold with boundary}, \emph{Commun. Part. Diff. Eq.} {\bf 15} (1990) 245.

\bibitem{MossPoletti}
I. Moss and S. Poletti, \emph{Boundary conditions for quantum cosmology}, \emph{Nucl. Phys. B} {\bf 341} (1990) 155.

\bibitem{Vassilevich1995}
D. V. Vassilevich, \emph{Vector fields on a disk with mixed boundary conditions}, \emph{J. Math. Phys.} {\bf 36} (1995) 3174 [\href{https://arxiv.org/abs/gr-qc/9404052}{\texttt{arXiv:gr-qc/9404052}}].

\bibitem{Vassilevich1998}
D. V. Vassilevich, \emph{The Faddeev–Popov trick in the presence of boundaries}, \emph{Phys. Lett. B} {\bf 421} (1998) 93 [\href{https://arxiv.org/abs/hep-th/9709182}{\texttt{arXiv:hep-th/9709182}}].

\bibitem{ChodosThorn}
A. Chodos and C. B. Thorn, \emph{Chiral invariance in a bag theory}, \emph{Phys. Rev. D} {\bf 12} (1975) 2733.

\bibitem{Yamagishi}
H. Yamagishi, \emph{Fermion-monopole system reexamined}, \emph{Phys. Rev. D} {\bf 27} (1983) 2383.

\bibitem{ZhangDietz}
W. Zhang and B. Dietz, \emph{Graphene billiards with fourfold symmetry}, \emph{Phys. Rev. Research} {\bf 5} (2023) 043028.

\bibitem{IvanovKurkovVassilevich}
A. V. Ivanov, M. A. Kurkov and D. V. Vassilevich, \emph{Heat kernel, spectral functions and anomalies in Weyl semimetals}, \emph{J. Phys. A: Math. Theor.} {\bf 55} (2022) 224004 [\href{https://arxiv.org/abs/2111.11493}{\texttt{arXiv:2111.11493}}].

\bibitem{EspositoKirsten}
G. Esposito and K. Kirsten, \emph{Chiral bag boundary conditions on the ball}, \emph{Phys. Rev. D} {\bf 66} (2002) 085014.

\bibitem{AtiyahPatodiSinger}
M. F. Atiyah, V. K. Patodi and I. M. Singer, \emph{Spectral asymmetry and Riemannian Geometry. I}, \emph{Math. Proc. Camb. Phil. Soc.} {\bf 77} (1975) 43.

\bibitem{Abrikosov}
A. A. Abrikosov Jr., \emph{Modified spectral boundary conditions in the bag model}, \emph{J. Phys. A: Math. Gen.} {\bf 39} (2006) 6109 [\href{https://arxiv.org/abs/hep-th/0512311}{\texttt{arXiv:hep-th/0512311}}].

\bibitem{Witten}
E. Witten, \emph{Fermion Path Integrals And Topological Phases}, \emph{Rev. Mod. Phys.} {\bf 88} (2016) 035001 [\href{https://arxiv.org/abs/1508.04715}{\texttt{arXiv:1508.04715}}].

\bibitem{DabholkarJainRudra}
A. Dabholkar, D. Jain and A. Rudra, \emph{APS $\eta$-invariant, path integrals, and mock modularity}, \emph{JHEP} {\bf 11} (2019) 080 [\href{https://arxiv.org/abs/1905.05207}{\texttt{arXiv:1905.05207}}].

\bibitem{FukayaFurutaMatsuoOnogiYamaguchi}
H. Fukaya, M. Furuta, S. Matsuo, T. Onogi and S. Yamaguchi, \emph{The Atiyah–Patodi–Singer Index and Domain-Wall Fermion Dirac Operators}, \emph{Commun. Math. Phys.} {\bf 320} (2020) 1295 [\href{https://arxiv.org/abs/1910.01987}{\texttt{arXiv:1910.01987}}].

\bibitem{KobayashiYonekura}
S. K. Kobayashi and K. Yonekura, \emph{Atiyah-Patodi-Singer index theorem from axial anomaly}, \emph{Prog. Theor. Exp. Phys.} {\bf 2021} (2021) 073B01 [\href{https://arxiv.org/abs/2103.10654}{\texttt{arXiv:2103.10654}}].

\bibitem{AsoreyBalachandranPerezPardo}
M. Asorey, A. P. Balachandran, J. M. Pérez-Pardo, \emph{Edge States: Topological Insulators, Superconductors and QCD Chiral Bags}, \emph{JHEP} {\bf 12} (2013) 073 [\href{https://arxiv.org/abs/1308.5635}{\texttt{arXiv:1308.5635}}].

\bibitem{DowkerGilkeyKirsten}
S. Dowker, P. B. Gilkey and K. Kirsten, \emph{Heat asymptotics with spectral boundary conditions}, \emph{Contemp. Math.} {\bf 242} (1999) 107 [\href{https://arxiv.org/abs/hep-th/0004020}{\texttt{arXiv:hep-th/0004020}}].

\bibitem{GilkeyKirsten}
P. B. Gilkey and K. Kirsten, \emph{Heat asymptotics with spectral boundary conditions II}, \emph{Proc. Roy. Soc. Edinburgh A} {\bf 113} (2003) 333 [\href{https://arxiv.org/abs/math-ph/0007015}{\texttt{arXiv:math-ph/0007015}}].

\bibitem{FoscoSilva}
C. D. Fosco and A. Silva, \emph{Chiral anomaly, induced current, and vacuum polarization tensor for a Dirac field in the presence of a defect}, \emph{Phys. Lett. B} {\bf 822} (2021) 136659 [\href{https://arxiv.org/abs/2108.00902}{\texttt{arXiv:2108.00902}}].

\bibitem{SundbergJaffe}
P. Sundberg and R. L. Jaffe, \emph{The Casimir effect for fermions in one dimension}, \emph{Ann. Phys.} {\bf 309} (2004) 442 [\href{https://arxiv.org/abs/hep-th/0308010}{\texttt{arXiv:hep-th/0308010}}].

\bibitem{GiesKlingmuller}
H. Gies and K. Klingmüller, \emph{Worldline algorithms for Casimir configurations}, \emph{Phys. Rev. D} {\bf 74} (2006) 045002 [\href{https://arxiv.org/abs/quant-ph/0605141}{\texttt{arXiv:quant-ph/0605141}}].

\bibitem{EdwardsGerberSchubertTrejoWeber}
J. P. Edwards, U. Gerber, C. Schubert, M. A. Trejo and A. Weber, \emph{Integral transforms of the quantum mechanical path integral: Hit function and path-averaged potential}, \emph{Phys. Rev. E} {\bf 97} (2018) 042114 [\href{https://arxiv.org/abs/1709.04984}{\texttt{arXiv:1709.04984}}].

\bibitem{EdwardsGerberSchubertTrejoTsiftsiWeber}
J. P. Edwards, U. Gerber, C. Schubert, M. A. Trejo, T. Tsiftsi and A. Weber, \emph{Applications of the worldline Monte Carlo formalism in quantum mechanics}, \emph{Ann. Phys.} {\bf 411} (2019) 167966 [\href{https://arxiv.org/abs/1903.00536}{\texttt{arXiv:1903.00536}}].






\end{thebibliography}
\end{document}